\documentclass[a4paper,UKenglish,cleveref, autoref, thm-restate]{oasics-v2021}
\pdfoutput=1 %ensure pdflatex processing
\graphicspath{{./figures/}}

\usepackage{xspace}
\usepackage{booktabs}
\usepackage{xurl}
\newcommand{\attrition}{24\xspace}
\newcommand{\trafficFeeds}{378\xspace}
\newcommand{\ixpLevelTPs}{352\xspace} % single
\newcommand{\fedLevelTPs}{26\xspace} % federated
\newcommand{\pdbInitialIXPs}{1065\xspace}
\newcommand{\noData}{586\xspace}
\newcommand{\noDataYesWebsite}{420\xspace}

\usepackage{tikz}
\usepackage{pgfplots}
\usepgfplotslibrary{fillbetween, groupplots}
\pgfplotsset{compat=1.18}

\usetikzlibrary{external, positioning, calc, patterns}

\definecolor{NorthAmerica}{HTML}{56B4E9} % North America – sky‑blue
\definecolor{SouthAmerica}{HTML}{009E73} % South America – bluish‑green
\definecolor{Europe}{HTML}{CC79A7}       % Europe – reddish‑purple
\definecolor{MiddleEast}{HTML}{F0E442}   % Middle East – yellow (desert sand)
\definecolor{Australia}{HTML}{E69F00}    % Australia – orange (outback ochre)
\definecolor{AsiaPacific}{HTML}{0072B2}  % Asia‑Pacific – blue (Pacific Ocean)
\definecolor{Africa}{HTML}{D55E00}       % Africa – vermilion (savanna earth)

\definecolor{Globe}{HTML}{FFDCB8} % Globe – sage-green (RGB average of all regions)
\definecolor{GlobeDark}{HTML}{8B976C} % Globe – sage-green (RGB average of all regions)

\definecolor{SinglePoint}{HTML}{3C4043}   % dark neutral, used for single data points
\definecolor{AggregatedPoint}{HTML}{BDBDBD} % light neutral, used for aggregated data points

\definecolor{Monday}{HTML}{D55E00}   % Vermilion
\definecolor{Tuesday}{HTML}{E69F00}  % Orange
\definecolor{Wednesday}{HTML}{F0E442}% Yellow
\definecolor{Thursday}{HTML}{009E73}% Bluish Green
\definecolor{Friday}{HTML}{56B4E9}   % Sky Blue
\definecolor{Saturday}{HTML}{0072B2} % Blue
\definecolor{Sunday}{HTML}{CC79A7}   % Reddish Purple
\title{\texorpdfstring{Five Blind Men and the Internet:\newline Towards an Understanding of Internet Traffic}{Five Blind Men and the Internet: Towards an Understanding of Internet Traffic}}

\titlerunning{Five Blind Men and the Internet: Towards an Understanding of Internet Traffic}

\author{Ege Cem Kirci}{Networked Systems Group, ETH Zürich, Switzerland}{ekirci@ethz.ch}{https://orcid.org/0000-0002-7014-1971}{}

\author{Ayush Mishra}{Networked Systems Group, ETH Zürich, Switzerland}{aymishra@ethz.ch}{https://orcid.org/0000-0001-9723-4513}{}

\author{Laurent Vanbever}{Networked Systems Group, ETH Zürich, Switzerland}{lvanbever@ethz.ch}{https://orcid.org/0000-0003-1455-4381}{}

\authorrunning{E. C. Kirci, A. Mishra, L. Vanbever} %TODO mandatory. First: Use abbreviated first/middle names. Second (only in severe cases): Use first author plus 'et al.'

\Copyright{Ege Cem Kirci, Ayush Mishra, Laurent Vanbever} %TODO mandatory, please use full first names. LIPIcs license is "CC-BY";  http://creativecommons.org/licenses/by/3.0/


\category{} %optional, e.g. invited paper

\relatedversion{} %optional, e.g. full version hosted on arXiv, HAL, or other respository/website
\acknowledgements{} %optional
\nolinenumbers %uncomment to disable line numbering

\EventEditors{Katerina J. Argyraki and Aurojit Panda}
\EventNoEds{2}
\EventLongTitle{1st New Ideas in Networked Systems (NINeS 2026)}
\EventShortTitle{NINeS 2026}
\EventAcronym{NINeS}
\EventYear{2026}
\EventDate{February 10, 2026}
\EventLocation{Virtual Conference}
\EventLogo{}
\SeriesVolume{139}
\ArticleNo{25}
\begin{document}

\maketitle

\begin{abstract}
Our current view of traffic on the Internet---the world's largest and most pervasive network---comes from a variety of perspectives, each with its own blind spots and biases. In this paper, we make the case for using publicly available Internet exchange point (IXP) statistics as a complementary vantage point. While IXP data has its own limitations, it is fine-grained, accessible, and independently verifiable---offering a distinct perspective on Internet usage patterns. We present results from a two-year study (2023--2024) of 472 IXPs worldwide, capturing approximately 300~Tbps of peak daily aggregate traffic by late 2024. Over this period, aggregate IXP traffic increased by 49.2\% (24.5\% annualized), with regionally distinct diurnal patterns and event-driven anomalies. These results provide an accessible framework for researchers and operators to study the Internet's evolving ecosystem from an IXP-based perspective, and lay the groundwork for systematic, global-scale detection of network anomalies and outages.
\end{abstract}

\section{Introduction}\label{sec:introduction}

Many (if not most) network decisions---be it about topology design, capacity planning, routing strategies, or monitoring---depend on the characteristics of carried traffic, such as total volume, expected growth, and inherent variability or {\em shape}.

This dependency holds at both the ``micro'' level, within a single Autonomous System~(AS), and the ``macro'' level, the Internet at large.
Internet-wide traffic characteristics directly impact CDN caching policies~\cite{bottgerOpenConnectEverywhere2018}, peering and deployment decisions~\cite{10.1145/3517745.3561462}, network economics~\cite{brown2024rcs}, and the scalability requirements of in-network algorithms~\cite{10.1145/3230543.3230544}.
For example, the surge of over 30\% in global Internet traffic during COVID-19 lockdowns prompted Netflix and YouTube to temporarily reduce video quality in Europe to ease network strain~\cite{goldNetflixYouTubeAre2020}.

\begin{figure}[t!]
	\begin{center}
		\includegraphics[width=0.45\columnwidth]{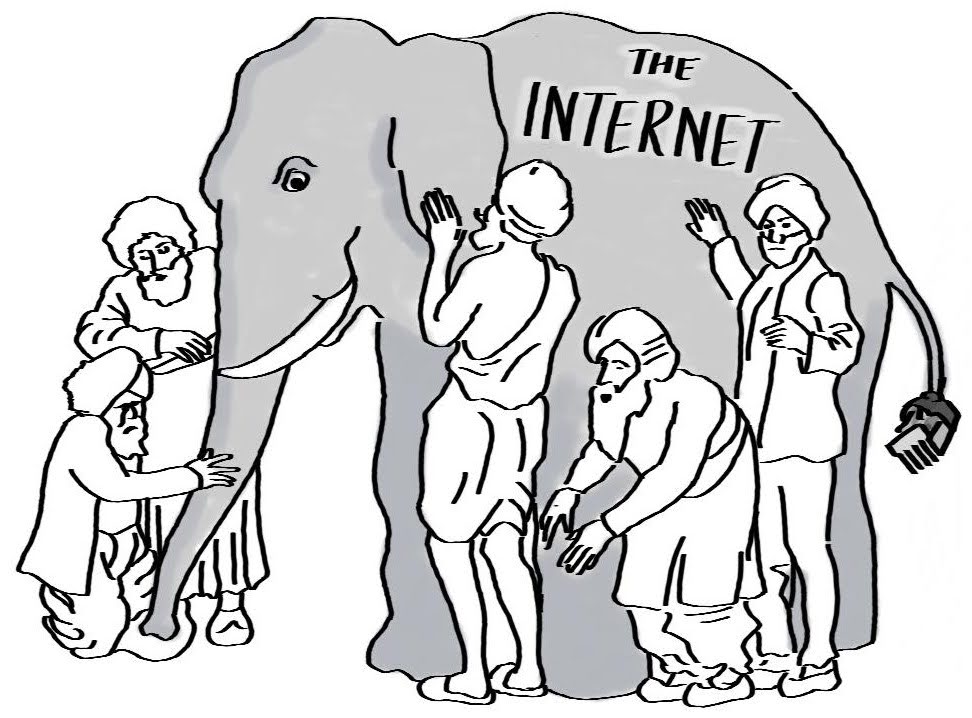}
		\caption{As in the parable of the Five Blind Men and the Elephant, each of the vantages for studying the Internet touches different parts of it---not the whole---and carries its own blind spots and biases.}
		\label{fig:elephant}
	\end{center}
\end{figure}

Yet, while tracking traffic at the level of a single AS is (relatively) straightforward, doing so at Internet scale is challenging.
The Internet is massive, diverse, and federated, which makes truly unbiased, global observability hard.
As in the parable of {\em The Five Blind Men and the Elephant}, illustrated in \cref{fig:elephant}, each vantage point touches only part of the Internet, and our understanding of Internet traffic's size, shape, and form is inevitably shaped by these biases.
At a high level, three primary sources inform our current picture of Internet traffic, each with its own inherent limitations:

First, measurements by major platforms serving users worldwide capture a {\em global} view of traffic patterns: for instance, studies from Facebook~\cite{bottgerHowInternetReacted2020,CANDELA2020107495,10.1145/3419394.3423658} observed post-COVID-19 traffic growth and shifting usage patterns.
While global in reach, such measurements are necessarily biased toward the specific applications and services these platforms provide~\cite{10.1145/3098822.3098853,10.1145/3452296.3472928}.

Second, measurements by major transit networks capture an {\em application-agnostic} view but typically only at local or regional scales, introducing geographical bias~\cite{10.1007/978-3-319-30505-9_25,10.1145/1851182.1851194}.

Finally, industry reports from companies such as Sandvine~\cite{marwaha_gipr_2024} and Cisco~\cite{cisco_air_2018_2023} aim to provide a global picture.
However, these reports are coarse-grained, often forecast-based rather than measured, rely on undisclosed or unverifiable assumptions, and are infrequent (typically annual).
They capture overall growth and volume but lack the granularity needed to study usage over shorter time horizons.

\paragraph*{A sixth Blind Man.} A representative view of the Internet emerges only by combining \emph{diverse}, \emph{high-quality}, and---crucially---\emph{complementary} sources. We argue that publicly available IXP traffic statistics provide such a complementary vantage: IXP traffic is broadly application-agnostic, more globally distributed, and easily accessible and verifiable. Like the other blind men, our vantage has its own blind spots---but it touches a different part of the elephant.

Prior work shows that IXPs capture a diverse range of Internet traffic, both in coverage~\cite{chatzisBenefitsUsingLarge2013} and application diversity~\cite{chatzisThereMoreIXPs2013, bottgerOpenConnectEverywhere2018}, and that IXPs play a significant role in how traffic is routed~\cite{richterPeeringPeeringsRole2014}.
As shown in \S\ref{sec:dataset-overview}, these statistics are available from diverse locations worldwide, updated frequently (typically every 5 minutes), and publicly accessible.\footnote{Since IXPs have an economic incentive to publish live traffic statistics to attract new members, it is likely that this information will remain available.}

While IXP traffic is not a precise estimator of the {\em total} amount of Internet traffic, with sufficient coverage it is rich enough to capture usage patterns and even distinguish atypical behavior during significant global events.
IXPs also carry substantial traffic in often under-studied regions such as Africa~\cite{10.1145/3646547.3689679}, improving visibility.

We present a two-year measurement study of traffic exchanged by IXPs worldwide from January 2023 to December 2024.
To collect these statistics, we built a collection framework that crawls IXP websites that make traffic statistics available online.
Overall, we tracked traffic at 472 IXPs worldwide.
These IXPs account for approximately 87\% of the total port capacity across the \pdbInitialIXPs IXPs listed in PeeringDB, a canonical registry of IXPs~\cite{PeeringDB}.
For these IXPs, we collected traffic volumes at 5-minute intervals throughout the entire study period.
The result is a rich dataset that supports inferences about IXP traffic growth and region-specific usage patterns---and, crucially, provides the validated baseline necessary for systematic detection of anomalies and outages at global scale.

\vspace{1em}
\noindent We make the following contributions and observations:

\begin{itemize}
	\item We built a reliable collection framework that gathers traffic volumes from 472 IXPs worldwide using API calls, HTML parsing, and optical character recognition (OCR).
	\item Average daily aggregate IXP traffic in our dataset rose by 49.2\% over two years (from 138~Tbps to 200~Tbps). The growth trends mirror recent Internet-traffic approximations by Cloudflare~\cite{cloudflare_year_in_review_2023,cloudflare_year_in_review_2024}, suggesting that IXP traffic is a strong proxy for studying Internet traffic growth. In some respects, the blind men agree.
	\item Yet their different perspectives also reveal divergences: our IXP data captures an increase during the 2024 Olympics Opening Ceremony, whereas Cloudflare observed a dip at the same time~\cite{tomeParis2024Olympics2024}.
	\item Thanks to our global coverage and fine-grained measurements, we observe distinct regional usage patterns, likely driven by differences in lifestyle and work schedules~(\S\ref{sec:usage-patterns}).
	\item Our framework also observes shifts from typical usage in response to global events such as the Olympics, national elections, e-commerce sales, and releases of popular game updates~(e.g. Fortnite).
\end{itemize}

\noindent To the best of our knowledge, this is the largest publicly-derived census of IXP traffic to date, surpassing prior work in scale, transparency, and duration.
This paper establishes the foundation---a validated, representative vantage point---upon which systematic global event detection can be built.
We are open-sourcing our dataset for further research, and our measurement framework continues to collect IXP traffic statistics.

\section{Background and Motivation}
\label{sec:background}

IXP data has been used to study the Internet in the past~\cite{10.1145/2602204.2602208,10.1145/3276799.3276801}.
In this section, we review prior studies, their insights, and some of their limitations, and use these observations to later motivate our framework's design.

\subsection{Prior IXP studies}

Prior work on traffic volumes exchanged at IXPs has typically relied on close collaboration with large exchanges such as AMS-IX and DE-CIX~\cite{bottger2018shaping,10.1145/2382016.2382018,10.1145/2342356.2342393}.
These studies showed that peak volumes at individual exchanges can reach tens of Tbps~\cite{10.1145/2342356.2342393,10.1145/2382016.2382018,10.1007/978-3-319-30505-9_25} and, thanks to fine-grained measurements, revealed distinct diurnal usage patterns in their regions.
Using anonymized sFlow records, they also quantified protocol and application-specific contributions to traffic volumes~\cite{10.1145/637201.637213,10.1145/1851182.1851194,10.1145/3544216.3544268}; for example, Feldmann et al.\ observed increased streaming and collaborative-tool usage during COVID-19~\cite{10.1145/3419394.3423658}.
Overall, this line of work demonstrates that IXP traffic is diverse, representative, and central to Internet connectivity~\cite{chatzisBenefitsUsingLarge2013,bottgerOpenConnectEverywhere2018}.

However, such studies have key limitations: because access often depends on operator agreements, vantage points are concentrated in a few regions, datasets are not easily reproduced or extended by others, and achieving global, longitudinal coverage without proprietary access is difficult. These constraints motivate a complementary approach.

\subsection{Observing global usage patterns}

Many IXPs publicly advertise aggregate traffic statistics to attract members. These feeds are mostly volumetric time series updated frequently (typically every 5 minutes) and are available across regions.
Because they are public, they are accessible and independently verifiable. Aggregating them can yield a broad, high-resolution, application-agnostic view of Internet usage patterns.

The challenge is practical rather than conceptual: formats vary (JSON APIs, HTML pages, and image-based plots), some feeds are federated, and care is required to avoid double counting and to normalize units, time zones, and sampling intervals. These realities shaped our framework's design (\S\ref{sec:methodology}).

It is important to note that we do not claim to see {\em all}, or even the {\em most significant} share of Internet traffic with this vantage. Like the other blind men, our vantage has its own blind spots (discussed in \S\ref{sec:discussion}).
Our goal is to capture the {\em shape} of Internet usage---its daily/weekly cycles, regional differences, and event-driven anomalies---rather than its absolute total.
As we will show in \S\ref{sec:usage-patterns}, with sufficient coverage this vantage consistently captures characteristic regional habits and identifies atypical behavior in response to global events.

\section{Methodology}\label{sec:methodology}

Our objective is to build a comprehensive, longitudinal dataset of traffic volumes exchanged at IXPs worldwide.
To this end, we continuously collect publicly available traffic profiles from all IXPs listed in PeeringDB as frequently as they are updated.
Achieving this goal requires overcoming key challenges in \emph{(i)}~data discovery, \emph{(ii)}~extraction, and \emph{(iii)}~system reliability.
In this section, we describe these challenges and how our system addresses them.

\subsection{Identifying {\em traffic feeds}}\label{sec:identifying-public-traffic-feeds}

PeeringDB is a community-maintained database widely recognized as the authoritative source for information on IXPs.
It catalogs publicly registered IXPs, their geographic locations, region classifications, participating networks (ASNs), those networks' port capacities, and peering policies, serving as a valuable starting point for acquiring traffic volume data.
PeeringDB also often provides URLs for IXPs that publicly share traffic statistics; however, we found this field to be sparsely populated and frequently inaccurate in practice.
We therefore manually inspected the websites of all IXPs listed in PeeringDB. This multi-week effort yielded a curated list of URLs for publicly accessible traffic statistics pages, enabling continuous monitoring of IXP traffic volumes.

These traffic feeds provide time-series data on outbound and inbound traffic exchanged over an IXP's public peering fabric, typically at five-minute intervals,
consistent with the Open-IX OIX-1 standard~\cite{oix_about, openix_oix1_2016}.
Discussions with 20 IXP operators confirm this practice represents total \textit{public} peering volume and excludes private network interconnects (PNIs);
accordingly, our analysis covers only the public fabric and does not capture PNIs.
To avoid double-counting, we use inbound traffic, noting that at the fabric aggregate inbound and outbound volumes are equal (what enters must exit).
Where both series were available, we observed they were virtually identical, with small differences attributable to packet loss within the fabric and measurement/reporting noise.

\begin{figure}[t]
	\centering
	\includegraphics[width=0.6\columnwidth]{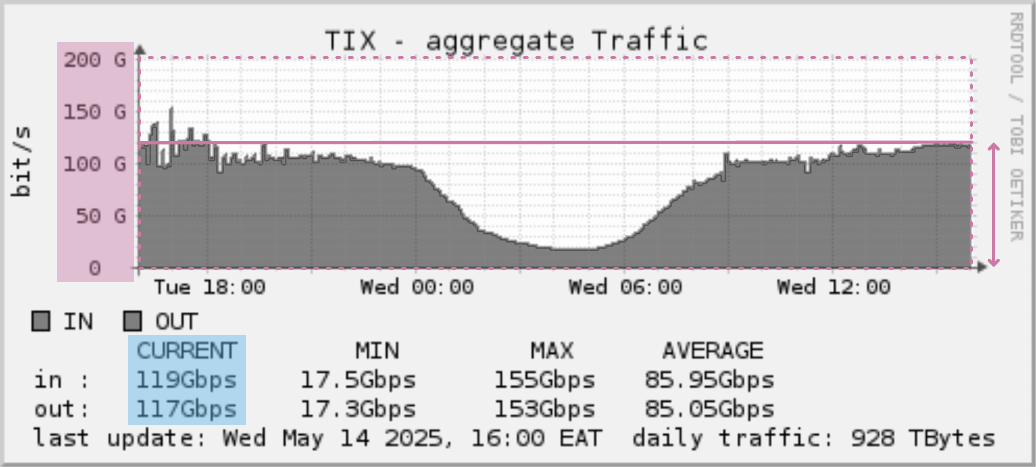}
	\caption{Example of OCR-based data extraction from graphical plots (Tanzania IX~\cite{tix_stats}). The red box and line illustrate digitization of plot axes and curve coordinates, while the blue box shows extraction of explicit textual data, such as ``Current: X~Gbps''.}\label{fig:ix_plot}
\end{figure}

\subsection{Extracting traffic data}\label{sec:extracting-traffic-data}

The one-time, labor-intensive process of identifying reliable traffic feeds is essential for enabling continuous and accurate extraction of IXP traffic volumes.
Upon establishing these feeds, our system methodically extracts traffic volumes from a diverse array of interfaces and formats.
Depending on the traffic feed, we employ one of the following three extraction methods:

\begin{description}
	\item[API-based.] For some IXPs, we directly obtain their traffic volume data in JSON format at five-minute intervals via IXP Manager~\cite{ixp_manager}.
	\item[HTML-based.] For IXPs not interfacing with IXP Manager, some embed traffic volume data within webpage HTML or accompanying JSON files. We develop specialized parsers using regular expressions specifically crafted for each individual IXP to extract this data.
	\item[OCR-based.] For remaining IXPs, we extract traffic data from graphical plots within images. We use the Tesseract OCR engine \cite{tesseract_ocr_5_5_0} with two complementary techniques: digitizing plot axes and curve coordinates, and extracting explicitly rendered text values (e.g.\ ``Current Traffic: X~Gbps''), when available (see \cref{fig:ix_plot}).
\end{description}

While API and HTML methods are generally stable once configured, OCR is inherently more fragile.
Potential errors include misinterpreting digits or missing decimal points.
To mitigate these, we implement a cross-check: when textual values are available on the plot, we compare them with the values derived from digitizing the plot curve.
If only plot digitization is possible, accuracy can be affected by image resolution.
We evaluated the accuracy of the OCR approach on a test set of images containing both plots and explicit text values.
The results revealed a maximum average deviation of 2.4\% between the OCR-extracted text values and the values derived from plot digitization, confirming the method's viability for our purposes.

The distribution of the \trafficFeeds traffic feeds in our study across extraction methods is relatively balanced: 146 rely on OCR, 126 use IXP Manager APIs, and 106 employ HTML parsing.
However, the share of traffic volume per method is skewed by operator size.
Feeds accessed via the standardized IXP Manager API contribute only 7.2\% of the total observed traffic volume,
whereas OCR-based and HTML-parsed sources---typically bespoke portals used by larger,
higher-traffic IXPs---account for roughly equal portions of the remainder.
This suggests that larger IXPs more often expose traffic via custom web interfaces rather than default presentation methods.

\subsection{Ensuring reliable collection}\label{sec:ensuring-reliable-collection}

Continuous, long-term data collection requires a robust and fault-tolerant infrastructure.
We designed and deployed a distributed scraping framework to automate data retrieval, validation, and storage.

Our system achieves fault tolerance primarily through redundancy.
The system maintains two nodes: the Main Node and the Backup Node.
Both nodes orchestrate scraping tasks based on configured frequencies, execute appropriate scrapers (API, HTML, OCR), perform initial data sanity checks, and transmit validated data to a separate Data Node hosting the database.
Careful job distribution prevents overload, ensuring smooth operation and consistent data quality.

While the Main Node actively performs these tasks, the Backup Node continuously monitors the Main Node's health by checking data freshness on the Data Node.
If the Main Node becomes unresponsive for longer than one polling interval plus a small grace window, the Backup Node automatically assumes scraping responsibilities.
This failover mechanism was triggered 57 times during the operational period, during scheduled maintenance and routine patching of the scraping hosts. The swift automated transition minimizes data gaps, ensuring the continuity of our longitudinal dataset.

The Data Node serves as the central repository for IXP traffic volume data, scraper configurations, historical PeeringDB snapshots~\cite{caida_peeringdb_2023_2024}, and operational logs.
It handles weekly data archiving and provides a Grafana-based interface for internal monitoring and visualization~\cite{grafana_12_0_0}.
Data storage employs a TimescaleDB database (a time-series optimized PostgreSQL extension)~\cite{timescaledb_2_19_3}.
This combination of redundancy, proactive monitoring, and specialized database architecture ensures robust, reliable, and scalable long-term data collection.

\section{Dataset}\label{sec:dataset-overview}

Our system enabled systematic collection of publicly available traffic statistics for IXPs worldwide over two years.
Of the~\pdbInitialIXPs IXPs listed in PeeringDB as of January 2023 (the start of our longitudinal study), we successfully collected continuous traffic data for 472 IXPs.
Although these represent slightly less than half of all IXPs listed in PeeringDB, they account for 87\% of the total port capacity of all IXPs.

\noindent The 472 IXPs analyzed derive their data from two types of traffic feeds:

\begin{description}
	\item[Single traffic feeds.] For 352 of the analyzed IXPs, we identified individual traffic feeds providing a one-to-one mapping between an IXP and its traffic data.
	\item[Federated traffic feeds.] For the remaining 120 IXPs, we identified traffic feeds that aggregate traffic for multiple IXPs managed by a single organization (e.g.\ Equinix). Consequently, these 120 IXPs were represented by 26 aggregate traffic feeds.
\end{description}

\noindent Thus, the 472 IXPs analyzed were represented by 378 traffic feeds, comprising single and federated feeds.

\paragraph*{Missed IXPs.} Of the \noData IXPs lacking traffic feeds, 166 had no identifiable websites, leaving their operational status uncertain.
For the remaining \noDataYesWebsite IXPs with websites, we found no publicly accessible traffic statistics.
Our classification relied on best-effort assessment and might contain errors, as some IXPs could host data in non-standard locations or behind logins.
Nevertheless, these missed IXPs, including those excluded due to attrition, constitute approximately 13\% of total IXP port capacity, suggesting that our dataset captures the majority of IXP traffic.

\begin{figure*}[t!]
	\begin{center}
		\input{figures/map/World}

\begin{tikzpicture}[scale=0.5]
    \WORLD[every state={draw=white, thick, fill=black!10}]
    
    % Load the data from CSVs
    \pgfplotstableread[col sep=comma]{figures/map/data/non_collected.csv}\noncollected
    \pgfplotstableread[col sep=comma]{figures/map/data/aggregated.csv}\aggregated
    \pgfplotstableread[col sep=comma]{figures/map/data/single.csv}\single
    
    % Plot non-collected points (black)
    \foreach \i in {1,...,612} {  % 614 lines total - 1 header
        \pgfmathtruncatemacro{\row}{\i-1}
        \pgfplotstablegetelem{\row}{x}\of\noncollected
        \pgfmathsetmacro{\x}{\pgfplotsretval}
        \pgfplotstablegetelem{\row}{y}\of\noncollected
        \pgfmathsetmacro{\y}{\pgfplotsretval}
        \ifx\x\empty\else
            \coordinate (Point\i) at (\x pt, \y pt);
            \fill[black] (Point\i) circle (1.75pt);
        \fi
    }
    
    % Plot aggregated points (orange)
    \foreach \i in {1,...,120} {  % 120 lines total - 1 header
        \pgfmathtruncatemacro{\row}{\i-1}
        \pgfplotstablegetelem{\row}{x}\of\aggregated
        \pgfmathsetmacro{\x}{\pgfplotsretval}
        \pgfplotstablegetelem{\row}{y}\of\aggregated
        \pgfmathsetmacro{\y}{\pgfplotsretval}
        \ifx\x\empty\else
            \coordinate (Point\i) at (\x pt, \y pt);
            \fill[Africa] (Point\i) circle (1.75pt);
        \fi
    }
    
    % Plot single points (positive green)
    \foreach \i in {1,...,352} {  % 337 lines total - 1 header
        \pgfmathtruncatemacro{\row}{\i-1}
        \pgfplotstablegetelem{\row}{x}\of\single
        \pgfmathsetmacro{\x}{\pgfplotsretval}
        \pgfplotstablegetelem{\row}{y}\of\single
        \pgfmathsetmacro{\y}{\pgfplotsretval}
        \ifx\x\empty\else
            \coordinate (Point\i) at (\x pt, \y pt);
            \fill[SouthAmerica] (Point\i) circle (1.75pt);
        \fi
    }

    % Add legend using working syntax
    \coordinate (Legend1) at (30pt, -290pt);
    \coordinate (Legend2) at (30pt, -270pt);
    \coordinate (Legend3) at (30pt, -250pt);
    
    \fill[black] (Legend1) circle (3pt);
    \node[anchor=west] at (Legend1) {Missed or excluded};
    
    \fill[Africa] (Legend2) circle (3pt);
    \node[anchor=west] at (Legend2) {Federated};
    
    \fill[SouthAmerica] (Legend3) circle (3pt);
    \node[anchor=west] at (Legend3) {Single};
\end{tikzpicture}
		\caption{Map of all IXPs listed in PeeringDB as of January 2023, classified by collection status.}\label{fig:map}
	\end{center}
\end{figure*}

\paragraph*{Excluded IXPs.} Beyond the 472 IXPs with continuous data, \attrition traffic feeds initially collected ceased operation during the two-year study due to unreachable websites or discontinued endpoints.
Despite efforts to contact operators, the reasons remained unclear.
Notably, 16 defunct feeds originated from conflict-affected regions (Russia: 5; Ukraine: 11), suggesting geopolitical instability as a potential factor.
Additionally, two IXPs, LL-IX and PIT-Arica, were later removed from PeeringDB, confirming their closure.
Analysis of their final weeks showed stable traffic volumes (41.6~Gbps for LL-IX and 4.5~Gbps for PIT-Arica), indicating abrupt termination rather than gradual decline.
Because PeeringDB is voluntary and lacks strict retirement protocols, some listed IXPs may be defunct, though confirmation is unavailable.

\vspace{0.5em}
\noindent This attrition resulted in \trafficFeeds feeds providing continuous time-series data throughout 2023--2024.
These active feeds consist of \ixpLevelTPs single-IXP feeds and \fedLevelTPs federated feeds.
Note that, except for the Equinix\footnote{Equinix later began sharing individual traffic feeds; however, to maintain continuity in our study, we still treat the data as federated.} feed (covering 42 IXPs distributed globally), all other federated feeds include IXPs located within the same region and time zone, as classified by PeeringDB's region and city fields.
We summarize the global distribution of single-IXP, federated, and missed IXPs in~\cref{fig:map}, with attrited IXPs counted as missed.
The IXPs in this figure are positioned according to their geographic coordinates listed in PeeringDB.
For IXPs spanning multiple sites, such as several data centers within a single city, we place the marker at the median of those coordinates.

\subsection{Data quality, coverage, and biases}
\label{sec:geographic-coverage-and-inherent-biases}

Given the lack of formal documentation for most public IXP traffic statistics, we assessed data quality along two key dimensions:

\begin{description}
	\item[Resolution.] Most traffic feeds (88\%) report data at five-minute intervals, aligning with the MRTG/RRD tool default and the MRTG standard used by IXP Manager. Another 9\% report at ten-minute intervals, with the remainder at coarser intervals, 30 minutes being the coarsest we observed.
	\item[Timeliness.] Most traffic feeds provide near real-time data, but four exhibit notable delays. The Equinix federated feed (covering 42 IXPs globally) and three IX.br locations (Campo Grande, Cascavel, Boa Vista) are five-minute time series, yet values are published as batch backfills and at uneven intervals rather than continuously. Our inquiries regarding these delays remain unanswered.
\end{description}

\paragraph*{The uneven landscape of IXPs worldwide.}
IXP infrastructure is far from uniformly distributed.
As shown in~\cref{tab:regional-distribution}, Europe (34.9\%), North America (21.3\%), and Asia-Pacific (21.1\%) host the majority (77.3\%) of the world's IXPs.
In contrast, Africa and the Middle East account for only 6.7\% and 2.2\%, respectively---an imbalance visible in \cref{fig:map}.

\begin{table}[t!]
	\small
	\centering
	\caption{Regional distribution and coverage of IXPs.}\label{tab:regional-distribution}
	\begin{tabular}{@{}lrrrr@{}}
		\toprule
		\textbf{Region} & \textbf{IXPs (\#)} & \textbf{Share (\%)} & \textbf{Collected (\#)} & \textbf{Coverage (\%)}\\
		\midrule
		Africa                     &  72 &  6.7 &  38 & 52.8\\
		Asia-Pacific              & 225 & 21.1 &  65 & 28.9\\
		Australia                  &  54 &  5.1 &  33 & 61.1\\
		Europe                     & 372 & 34.9 & 185 & 49.7\\
		Middle East                &  23 &  2.2 &  10 & 43.5\\
		North America                 & 227 & 21.3 &  88 & 38.8\\
		South America                 & 92 &  8.7 &  53 & 57.6\\
		\midrule
		Global total               & \pdbInitialIXPs & 100 & 472 & 44.4\\
		\bottomrule
	\end{tabular}
\end{table}

This imbalance intensifies at finer granularities.
The United States leads all countries with 191 registered IXPs, while approximately half of all countries operate just one, and 50 nations have none.
Indonesia, Brazil, Australia, Germany, India, and Russia each host over 40 IXPs.
At the metropolitan scale, connectivity concentrates dramatically: a handful of hyper-connected cities---Jakarta (15 IXPs), Amsterdam (14), Frankfurt (13), London (12), and Singapore (11)---boast double-digit exchanges, whereas the median city with an IXP operates just one.

\paragraph*{Measurement coverage and blind spots.}
Beyond regional counts, our measurement coverage also varies significantly (see \cref{tab:regional-distribution}).
We achieved the highest coverage in Australia (over 60\%)\footnote{The Australia region refers to Australia and New Zealand.}, South America (over 50\%), and Africa (over 50\%).
All regions except North America and Asia-Pacific exceed 40\% coverage.
Coverage is notably lower in North America (38.8\%) and especially in Asia-Pacific (28.9\%).
Overall, our dataset encompasses traffic data for 44.4\% of all IXPs listed in PeeringDB.

\begin{figure}[t!]
	\begin{center}
		\begin{tikzpicture}
    \pgfplotsset{set layers}
    \begin{scope}[line join=round]
        \begin{axis}[
                ymin=0, ymax=180,
                tick align=outside,
                tick pos=left,
                axis line style=thick,
                axis x line*=bottom,
                x axis line style={},
                y axis line style={->, >=latex},
                axis y line*=left,
                major tick length=0.2cm,
                xtick style={color=black, thick},
                ytick style={color=black, thick},
                ylabel={\small Average traffic (Tbps)},
                xticklabel style={font=\small},
                yticklabel style={font=\small},
                height=4.5cm,width=0.90\linewidth,
                legend style={
                        draw=none,
                        font=\small,
                        legend columns=4,
                        column sep=0.1cm,
                        row sep=0.05cm,
                        anchor=center,
                        at={(0.42, 1.25)},
                        legend cell align={left},
                        cells={fill=none},
                        fill=none
                    },
                ymin=0, ymax=180,
                ytick={0, 40, 80, 120, 160},
                yticklabels={0, 40, 80, 120, 160},
                xmin=0, xmax=733,
                xtick={1, 90, 180, 270, 360, 450, 540, 630, 720},
                xticklabels={Jan, Apr, Jul, Oct, Jan, Apr, Jul, Oct, Dec},
                clip = false,
                set layers=standard
            ]

            \path[name path=x_axis] (axis cs:0,0) -- (axis cs:733,0);

            \addlegendimage{area legend, fill=Africa, draw=Africa}
            \addlegendentry {Africa}
            \addlegendimage{area legend, fill=AsiaPacific, draw=AsiaPacific}
            \addlegendentry {Asia-Pacific}
            \addlegendimage{area legend, fill=Australia, draw=Australia}
            \addlegendentry {Australia}
            \addlegendimage{area legend, fill=Europe, draw=Europe}
            \addlegendentry {Europe}
            \addlegendimage{area legend, fill=MiddleEast, draw=MiddleEast}
            \addlegendentry {Middle East}
            \addlegendimage{area legend, fill=NorthAmerica, draw=NorthAmerica}
            \addlegendentry {North America}
            % \addlegendimage{empty legend}
            % \addlegendentry {}
            % \addlegendimage{empty legend}
            % \addlegendentry {}
            \addlegendimage{area legend, fill=SouthAmerica, draw=SouthAmerica}
            \addlegendentry {South America}

            \addplot+[name path=middle_east, no marks, thick, color=MiddleEast] table[col sep=comma, x=day_number, y=Middle East] {figures/stacked_traffic/data/regional_stacked_traffic.csv};
            \addplot+[name path=australia, no marks, thick, color=Australia] table[col sep=comma, x=day_number, y=Australia] {figures/stacked_traffic/data/regional_stacked_traffic.csv};
            \addplot+[name path=africa, no marks, thick, color=Africa] table[col sep=comma, x=day_number, y=Africa] {figures/stacked_traffic/data/regional_stacked_traffic.csv};
            \addplot+[name path=north_america, no marks, thick, color=NorthAmerica] table[col sep=comma, x=day_number, y=North America] {figures/stacked_traffic/data/regional_stacked_traffic.csv};
            \addplot+[name path=asia_pacific, no marks, thick, color=AsiaPacific] table[col sep=comma, x=day_number, y=Asia Pacific] {figures/stacked_traffic/data/regional_stacked_traffic.csv};
            \addplot+[name path=south_america, no marks, thick, color=SouthAmerica] table[col sep=comma, x=day_number, y=South America] {figures/stacked_traffic/data/regional_stacked_traffic.csv};
            \addplot+[name path=europe, no marks, thick, color=Europe] table[col sep=comma, x=day_number, y=Europe] {figures/stacked_traffic/data/regional_stacked_traffic.csv};

            \addplot+[color=MiddleEast] fill between[of=middle_east and australia, on layer=axis grid];
            \addplot+[color=Australia] fill between[of=australia and africa, on layer=axis grid];
            \addplot+[color=Africa] fill between[of=africa and north_america, on layer=axis grid];
            \addplot+[color=NorthAmerica] fill between[of=north_america and asia_pacific, on layer=axis grid];
            \addplot+[color=AsiaPacific] fill between[of=asia_pacific and south_america, on layer=axis grid];
            \addplot+[color=SouthAmerica] fill between[of=south_america and europe, on layer=axis grid];
            \addplot+[color=Europe] fill between[of=europe and x_axis, on layer=axis grid];

            %\node[left] at (axis description cs:0,1.05) {Tbps};
        \end{axis}
    \end{scope}
\end{tikzpicture}
		\caption{Stacked plot of daily mean traffic volume by region, with Europe contributing 49\% of observed traffic, based on PeeringDB's region classifications.}\label{fig:stacked_traffic}
	\end{center}
\end{figure}

The shortfall in Asia-Pacific is largely attributable to policy-driven opacity in two major economies: China contributes no public traffic feeds, and India provides data for only 23\% of its IXPs.
Attempts to access IXP websites in these countries often failed from our crawlers, suggesting potentially restrictive sharing practices or network filtering.
However, opacity is not solely an Asian phenomenon; for example, Fremont, California---a major U.S. technology hub---publishes data for only one of its ten registered IXPs.
Addressing these coverage gaps remains a key direction for future work.

\paragraph*{Traffic volume bias.}
Observed traffic shares vary across regions and reflect measurement coverage.
In~\cref{fig:stacked_traffic}, we present a stacked plot of average traffic volume by region, excluding Equinix due to its multi-continent footprint.\footnote{Equinix is likewise excluded from all further regional analyses in this paper.}

European IXPs account for approximately 70.2~Tbps, or roughly 49\% of the total average daily traffic observed (144~Tbps).
This dominance stems from market realities---Europe hosts some of the world's largest IXPs---and high measurement accessibility from major exchanges like DE-CIX Frankfurt, AMS-IX, LINX, and Netnod Stockholm.

North America and Asia-Pacific rank second and third by volume, respectively, but with lower shares than one might expect given their IXP counts, due to limited public data from many large exchanges and the exclusion of Equinix from regional analyses.
South America, with high coverage (57.6\%), contributes 29.0~Tbps (20\%).
Africa (3.6~Tbps, 2.5\%), the Middle East (1.3~Tbps, 1.0\%), and Australia (1.7~Tbps, 1.2\%) each contribute smaller shares, despite average or above-average coverage (\cref{tab:regional-distribution}).
Nevertheless, each of these regions surpasses the 1~Tbps daily average threshold, which we consider sufficient for meaningful volume-based analysis within these regions (see \S\ref{subsec:growth-regional-view}).

\subsection{Representativeness}\label{ssec:representativeness-via-port-capacity}

Given the evident geographic and traffic volume disparities, how representative is our dataset of IXP infrastructure?

As a measure of representativeness, we divided the total port capacities of all the IXPs we were successfully able to measure in a region by the total port capacities of {\em all} the IXPs in that region. We plot these ratios in~\cref{fig:scraped_capacity} for each of our seven regions. Since IXPs updated their port capacities during our measurement period, we plot these ratios over the entire measurement period.

As we can see, there is strong evidence of representativeness. Globally, our monitored IXPs account for 87\% of all publicly announced IXP port capacity listed in PeeringDB. This high capacity coverage remained remarkably stable over 2023--2024, decreasing by only 1.5 percentage points despite considerable growth in IXP traffic in this period. This high coverage persists regionally, with our framework observing more than 80\% available port capacity in all regions. Ranked by coverage, Australia leads (98\%), followed by South America (96\%) and Africa (92\%).

We note that the analysis presented in \cref{fig:scraped_capacity} exclude 76 IXPs (out of \pdbInitialIXPs) lacking port capacity data in PeeringDB during our study period; manual checks suggest these are typically small, geographically dispersed, or inactive IXPs, with likely negligible impact on aggregate results.

Also, because PeeringDB is self-reported, we assessed the freshness and internal consistency of the capacity data used for this analysis.
For all IXPs included in our capacity analysis, the port-capacity-related fields were updated at least once per year during 2023--2024, including IXPs without public traffic feeds, indicating that these records are actively maintained rather than stale.
As a plausibility check, where both traffic and capacity were available we verified that observed peaks did not exceed announced capacities; and we found no systematic contradictions.

\begin{figure}[t!]
	\begin{center}
		\begin{tikzpicture}
    \begin{axis}[
        tick align=outside,
        tick pos=left,
        axis line style=thick,
        axis x line*=bottom,
        x axis line style={},
        y axis line style={},
        axis y line*=left,
        major tick length=0.2cm,
        xtick style={color=black, thick},
        ytick style={color=black, thick},
        xticklabel style={font=\small},
        yticklabel style={font=\small},
        ylabel={Port capacity ratio},
        ymin=0.75, ymax=1,
        xmin=0, xmax=731,
        x tick label as interval=false,
        xtick={0, 90, 180, 270, 360, 450, 540, 630, 720},
        xticklabels={Jan, Apr, Jul, Oct, Jan, Apr, Jul, Oct, Dec},
        ytick={0.75, 0.8, 0.85, 0.9, 0.95, 1.0},
        yticklabels={0.75, 0.8, 0.85, 0.9, 0.95, 1.0},
        height=4.5cm,
        width=0.90\linewidth,
        clip=false,
        legend style={
            at={(0.42,1.25)}, 
            anchor=center, 
            draw=none, 
            legend columns=4, 
            column sep=0.1cm, 
            row sep=0.05cm, 
            font=\small, 
            legend cell align={left},
            legend image post style={line width=2pt}, 
            fill=none},
        legend cell align=left,
        set layers=standard
    ]
       
    \addplot[color={Africa}, thick] table[col sep=comma,x=day_index,y=ratio_africa] {figures/scraped_capacity/regional_scraped_ratio.csv};
    \addlegendentry{Africa}
 
    \addplot[color={AsiaPacific}, thick] table[col sep=comma,x=day_index,y=ratio_asia_pacific] {figures/scraped_capacity/regional_scraped_ratio.csv};
    \addlegendentry{Asia-Pacific}

    \addplot[color={Australia}, thick] table[col sep=comma,x=day_index,y=ratio_australia] {figures/scraped_capacity/regional_scraped_ratio.csv};
    \addlegendentry{Australia}
     
    \addplot[color={Europe}, thick] table[col sep=comma,x=day_index,y=ratio_europe] {figures/scraped_capacity/regional_scraped_ratio.csv};
    \addlegendentry{Europe}
    
    \addplot[color={MiddleEast}, thick] table[col sep=comma,x=day_index,y=ratio_middle_east] {figures/scraped_capacity/regional_scraped_ratio.csv};
    \addlegendentry{Middle East}
 
    \addplot[color={NorthAmerica}, thick] table[col sep=comma,x=day_index,y=ratio_north_america] {figures/scraped_capacity/regional_scraped_ratio.csv};
    \addlegendentry{North America}

    \addplot[color={SouthAmerica}, thick] table[col sep=comma,x=day_index,y=ratio_south_america] {figures/scraped_capacity/regional_scraped_ratio.csv};
    \addlegendentry{South America}

    \addplot[color={black}, thick] table[col sep=comma,x=day_index,y=ratio_global] {figures/scraped_capacity/regional_scraped_ratio.csv};
    \addlegendentry{Global}

    \end{axis}
\end{tikzpicture}
		\caption{Port capacity coverage of 472 monitored IXPs, representing 87\% of total IXP port capacity.}\label{fig:scraped_capacity}
	\end{center}
\end{figure}

\section{Traffic Growth}\label{sec:traffic-growth}

In this section, we examine overall IXP-traffic growth during our two-year measurement period and analyze how trends vary seasonally and across regions.

\subsection{Global traffic expansion}\label{subsec:global-growth-trajectory-and-volatility}

Between January 2023 and December 2024, aggregate IXP traffic exhibited substantial and consistent growth.
We illustrate this trend in \cref{fig:global_trends}, which sums the daily mean, peak ($95^{th}$\,percentile), and trough ($5^{th}$\,percentile) traffic volumes across all traffic profiles to provide a comprehensive view of aggregate IXP traffic volume.
The daily mean traffic rose from approximately 138~Tbps to 200~Tbps, a 49.2\% increase over the two-year period.
Peak and trough traffic experienced similar increases of 47.6\% and 54.3\%, respectively.

While substantial, growth was not uniform: average traffic increased by 23.4\% in 2023 but by a more moderate 16.9\% in 2024, based on a linear fit applied to each year.
These figures align closely with independent observations (Cloudflare reporting 25.0\% and 17.2\% global increases, respectively)~\cite{cloudflare_year_in_review_2023,cloudflare_year_in_review_2024}, suggesting that IXP traffic is a strong proxy for studying overall Internet-traffic growth.
Peak traffic followed a similar deceleration pattern (20.0\% in 2023 vs. 13.6\% in 2024), while trough traffic slowed less dramatically (27.7\% vs. 22.2\%).

Beyond growth rates, the day-to-day volatility of these metrics reveals distinct characteristics.
Peak traffic exhibits the largest absolute daily fluctuations, with typical swings of $\pm$6.0~Tbps (standard deviation).
Conversely, trough traffic is proportionally the most volatile, with daily changes averaging 5.1\% relative to its mean---more than double the relative volatility of peak (2.4\%) or average (2.0\%) traffic.
Average traffic is the most stable metric, with absolute fluctuations of $\pm$3.4~Tbps and the lowest relative volatility, making it ideal for tracking overall trends reliably.

\begin{figure}[t!]
	\begin{center}
		\begin{tikzpicture}
    \begin{axis}[
        tick align=outside,
        tick pos=left,
        axis line style=thick,
        axis x line*=bottom,
        x axis line style={},
        y axis line style={->, >=latex},
        axis y line*=left,
        major tick length=0.2cm,
        xtick style={color=black, thick},
        ytick style={color=black, thick},
        xticklabel style={font=\small},
        yticklabel style={font=\small},
        ylabel={Traffic (Tbps)},
        ymin=0, ymax=340,
        xmin=0, xmax=731,
        xtick={0, 90, 180, 270, 360, 450, 540, 630, 720},
        xticklabels={Jan, Apr, Jul, Oct, Jan, Apr, Jul, Oct, Dec},
        ytick={0, 100, 200, 300},
        yticklabels={0, 100, 200, 300},
        height=4.5cm,
        width=0.90\linewidth,
        clip=false,
        legend style={at={(0.5,1.15)}, anchor=north, draw=none, legend columns=3, column sep=0.15cm, font=\small, legend image post style={line width=2pt}},
        legend cell align=center,
        set layers=standard
    ]
    
    % Add background rectangles on the axis background layer
    \pgfplotsextra{
        \begin{pgfonlayer}{axis background}
            \fill[AggregatedPoint] (axis cs:0,0) rectangle (axis cs:180,310);
            \fill[AggregatedPoint] (axis cs:360,0) rectangle (axis cs:540,310);
        \end{pgfonlayer}
    }
    
    % Add the three lines
    \addplot[color={AsiaPacific}, thick] table[col sep=comma,x=day_number, y=total_95th_percentile_tb] {figures/global_trends/data.csv};
    \addlegendentry{95th \%}
    \addplot[color={SouthAmerica}, thick] table[col sep=comma,x=day_number, y=total_daily_traffic_tb] {figures/global_trends/data.csv};
    \addlegendentry{Mean}
    \addplot[color={Africa}, thick] table[col sep=comma,x=day_number, y=total_5th_percentile_tb] {figures/global_trends/data.csv};
    \addlegendentry{5th \%}
    
    \end{axis}
    \end{tikzpicture}
		\caption{Aggregate IXP traffic grew 49\% over two years (23.4\% in 2023, 16.9\% in 2024).}
        \label{fig:global_trends}
	\end{center}
\end{figure}

\subsection{Seasonal cycles and holiday impacts}\label{subsec:seasonal-cycles-and-holiday-impacts}

Traffic growth varies significantly within each year of the study period.
As evident in \cref{fig:global_trends}, aggregate traffic growth remains relatively stagnant in the first half of the year (shaded in gray) but accelerates in the second half.
For example, in 2023, average traffic grew by only 3.4\% in the first half of the year, compared to 14.3\% in the second half.

Regional analysis clarifies this trend. As illustrated in \cref{fig:trends_regions}, northern hemisphere regions closely follow the aggregate pattern, whereas South America exhibits the opposite trend.
For instance, in 2023, South America experienced accelerated traffic growth of 42.5\% in the first half of the year, compared to only 17.1\% in the second half.
Australia, another southern hemisphere region, showed balanced growth throughout the year.
A common pattern emerges: traffic growth tends to accelerate during cooler months of the year, possibly due to increased indoor online activity during autumn and winter.

\begin{figure}[t!]
	\begin{center}
		\begin{tikzpicture}
    \begin{axis}[
        tick align=outside,
        tick pos=left,
        axis line style=thick,
        axis x line*=bottom,
        x axis line style={},
        y axis line style={->, >=latex},
        axis y line*=left,
        major tick length=0.2cm,
        xtick style={color=black, thick},
        ytick style={color=black, thick},
        xticklabel style={font=\small},
        yticklabel style={font=\small},
        ylabel={Growth rate (\%)},
        ymin=-10, ymax=145,
        xmin=0, xmax=731,
        xtick={0, 90, 180, 270, 360, 450, 540, 630, 720},
        xticklabels={Jan, Apr, Jul, Oct, Jan, Apr, Jul, Oct, Dec},
        ytick={0, 25, 50, 75, 100, 125},
        yticklabels={0, 25, 50, 75, 100, 125},
        height=4.5cm,
        width=0.90\linewidth,
        clip=false,
        legend style={at={(0.42,1.25)}, anchor=center, draw=none, legend columns=4, column sep=0.1cm, row sep=0.05cm, font=\small, legend image post style={line width=2pt}, fill=none},
        legend cell align=left,
        set layers=standard
    ]
    
    % Add the region lines
    \addplot[color={Africa}, thick] table[col sep=comma,x=day,y=Africa] {figures/regions/regional_growth_data.csv};
    \addlegendentry{Africa}
    
    \addplot[color={AsiaPacific}, thick] table[col sep=comma,x=day,y={Asia Pacific}] {figures/regions/regional_growth_data.csv};
    \addlegendentry{Asia-Pacific}
    
    \addplot[color={Australia}, thick] table[col sep=comma,x=day,y=Australia] {figures/regions/regional_growth_data.csv};
    \addlegendentry{Australia}
    
    \addplot[color={Europe}, thick] table[col sep=comma,x=day,y=Europe] {figures/regions/regional_growth_data.csv};
    \addlegendentry{Europe}
    
    \addplot[color={MiddleEast}, thick] table[col sep=comma,x=day,y={Middle East}] {figures/regions/regional_growth_data.csv};
    \addlegendentry{Middle East}
    
    \addplot[color={NorthAmerica}, thick] table[col sep=comma,x=day,y={North America}] {figures/regions/regional_growth_data.csv};
    \addlegendentry{North America}

    \addplot[color={SouthAmerica}, thick] table[col sep=comma,x=day,y={South America}] {figures/regions/regional_growth_data.csv};
    \addlegendentry{South America}
    
    \end{axis}
\end{tikzpicture}
		\caption{Traffic surged substantially across all regions, with the Asia-Pacific region experiencing the greatest increase and Europe the most modest rise.}
		\label{fig:trends_regions}
	\end{center}
\end{figure}

Short-term slowdowns appear around specific holidays and events, including the Christmas period in Europe and North America, New Year's Eve/Day in North America, and Eid al-Fitr (April 21-23, 2023, and April 10-12, 2024), marking the end of Ramadan, in the Middle East.
These slowdowns likely reflect reduced business activity and shifts in user behavior during holidays.
Note that \cref{fig:global_trends} uses a seven-day rolling (weekly) window average for visual clarity, which smooths short-lived, extreme fluctuations; the unsmoothed series shows these holiday dips more prominently.

These examples demonstrate that, while broad seasonal cycles provide a baseline, specific cultural and regional events introduce predictable, short-term variations in aggregated IXP traffic data.
Understanding both large-scale seasonal waves and short-term, event-driven ripples is crucial for accurately interpreting traffic dynamics, a theme further explored in~\S\ref{sec:usage-patterns}.

\subsection{Regional divergence}\label{subsec:growth-regional-view}

Aggregate averages mask substantial regional heterogeneity (\cref{fig:trends_regions}).
Europe, the largest region by observed volume, shows the most modest growth (15.5\%) over two years, while Asia-Pacific (108\%), Middle East (82\%), North America (72\%), Africa (68\%), and South America (60\%) grow markedly faster, reflecting rapid development, successful interconnection initiatives, or strong demand growth.
Australia shows moderate growth (43\%).

Given the disparities in observed traffic volumes across regions~(\S\ref{sec:geographic-coverage-and-inherent-biases}), we investigated whether these different growth rates were genuine or artifacts of varying coverage and observed traffic volumes across regions.
To do so, we analyzed how quickly each region's aggregate growth pattern converges as more IXP data is included.
Starting with the IXP contributing the least traffic in a region and progressively adding larger ones by volume, we calculated the Pearson correlation (denoted $R$) between the two-year growth trend of this cumulative partial aggregate and the growth trend of all observed IXPs in that region.
The results are plotted in \cref{fig:growth_corr}.

For all regions except the Middle East and Australia, correlations increase steadily as more traffic is included, indicating high self-similarity within subsets of IXPs.
In the Middle East, Manama-IX (Bahrain), a major contributor to regional traffic, experienced an abrupt 50\% drop in average traffic (from 300~Gbps to 150~Gbps) around September 2023, likely due to a network disconnection or contract termination.
This sudden shift in a single IXP, diverging from smoother regional trends (yellow line in~\cref{fig:growth_corr}), delays cumulative correlation.
A similar, less pronounced event occurred in Australia with Megaport Brisbane in early January 2024, possibly due to a contract termination.
These disruptions, while skewing convergence metrics for the Middle East and Australia, reinforce confidence in the data's fidelity, suggesting that, absent such events, convergence would likely occur earlier.

\cref{tab:regional-growth-correlation} summarizes the observed-traffic share needed to achieve $R{>}0.90$ and $R{>}0.95$ with the final regional trend.
In most regions, growth trends stabilize remarkably quickly.
For instance, Asia-Pacific and North America reach $R{>}0.95$ with less than 10\% of observed traffic volume.
South America and Europe require 57.8\% and 56.8\%, respectively, while Africa and Australia achieve $R{=}0.95$ at 35.0\% and 48.1\%.
The Middle East, due to the Manama-IX disruption, requires 72.6\% for both thresholds.

\paragraph*{Takeaway:} the growth-rate signal converges quickly, so we do not need complete coverage to recover the regional trend; once a modest share of traffic is included (often $<\!10\%$), adding more IXPs primarily narrows uncertainty rather than changing the trajectory.

\vspace{0.5em}
\noindent Overall, the dominant growth signal emerges well before all data is aggregated, varying by region.
This internal consistency, combined with our high capacity coverage~(\S\ref{ssec:representativeness-via-port-capacity}), strongly supports the representativeness of our findings.
The convergence suggests that the observed growth rates and regional trajectories reported here accurately reflect the dominant dynamics of the broader regional ecosystems, rather than being artifacts of incomplete measurement.
This robustness is further explored from an infrastructure perspective in~\S\ref{sec:supplementary}.

\begin{figure}[t!]
	\begin{center}
		\begin{tikzpicture}
    \begin{axis}[
        tick align=outside,
        tick pos=left,
        axis line style=thick,
        axis x line*=bottom,
        x axis line style={},
        y axis line style={},
        axis y line*=left,
        major tick length=0.2cm,
        xtick style={color=black, thick},
        ytick style={color=black, thick},
        ylabel={Correlation},
        xlabel={Proportion of total traffic},
        ymin=0, ymax=1,
        xmin=0.01, xmax=1,
        xmode=log,
        xtick={0.01, 0.1, 1},
        xticklabels={1\%, 10\%, 100\%},
        ytick={-1, -0.5, 0, 0.5, 1},
        yticklabels={-1, -0.5, 0, 0.5, 1},
        height=4.0cm,
        width=0.90\linewidth,
        clip=false,
        set layers=standard,
        tick label style={font=\small},
        legend style={
            at={(0.42,1.25)}, 
            draw=none,
            anchor=center, 
            legend columns=4, 
            column sep=0.1cm, 
            row sep=0.05cm, 
            font=\small, 
            legend image post style={line width=2pt}, 
            legend cell align=left, 
            fill=none},
        legend cell align=left,
    ]
    
    \addplot[color={Africa}, thick, mark=*, mark size=1pt] table[col sep=comma,x=x,y=y] {figures/growth_corr/data/africa_cumulative_correlation.csv};
    \addlegendentry{Africa}
    
    \addplot[color={AsiaPacific}, thick, mark=*, mark size=1pt] table[col sep=comma,x=x,y=y] {figures/growth_corr/data/asia_pacific_cumulative_correlation.csv};
    \addlegendentry{Asia-Pacific}
    
    \addplot[color={Australia}, thick, mark=*, mark size=1pt] table[col sep=comma,x=x,y=y] {figures/growth_corr/data/australia_cumulative_correlation.csv};
    \addlegendentry{Australia}
    
    \addplot[color={Europe}, thick, mark=*, mark size=1pt] table[col sep=comma,x=x,y=y] {figures/growth_corr/data/europe_cumulative_correlation.csv};
    \addlegendentry{Europe}
    
    \addplot[color={MiddleEast}, thick, mark=*, mark size=1pt] table[col sep=comma,x=x,y=y] {figures/growth_corr/data/middle_east_cumulative_correlation.csv};
    \addlegendentry{Middle East}
    
    \addplot[color={NorthAmerica}, thick, mark=*, mark size=1pt] table[col sep=comma,x=x,y=y] {figures/growth_corr/data/north_america_cumulative_correlation.csv};
    \addlegendentry{North America}
    
    \addplot[color={SouthAmerica}, thick, mark=*, mark size=1pt] table[col sep=comma,x=x,y=y] {figures/growth_corr/data/south_america_cumulative_correlation.csv};
    \addlegendentry{South America}
    
    \end{axis}
\end{tikzpicture}
		\caption{Convergence of regional growth signals as more traffic is included. The Middle East shows irregular convergence due to a step change at a single IXP; other regions converge smoothly.}\label{fig:growth_corr}
	\end{center}
\end{figure}

\begin{table}[t!]
    \small
    \centering
    \caption{Regional growth trends stabilize rapidly, requiring only a fraction of observed traffic to achieve strong correlation with the final regional trend.}\label{tab:regional-growth-correlation}
\begin{tabular}{@{}lrr@{}}
	\toprule
	\textbf{Region}      & \textbf{\% Obs. traffic for R=0.90} & \textbf{\% Obs. traffic for R=0.95} \\
	\midrule
	Africa        &                          18.4\% &                         35.0\% \\
	Asia-Pacific  &                          0.2\% &                          0.3\% \\
	Australia     &                          8.3\% &                         48.1\% \\
	Europe        &                          1.9\% &                         56.8\% \\
	Middle East   &                         72.6\% &                         72.6\% \\
	North America &                          5.3\% &                          8.1\% \\
	South America &                         23.9\% &                         57.8\% \\
	\bottomrule
\end{tabular}
\end{table}

\begin{figure}[t!]
	\begin{center}
		\begin{tikzpicture}
	\begin{groupplot}[
			group style={
					group size=1 by 3,
					vertical sep=5pt, % further reduced separation between plots
				},
			width=0.90\linewidth,
			height=2.5cm,
			xlabel near ticks,                % x‑label tight to the axis
			ylabel near ticks,                % y‑label tight to the axis
			label style={font=\small},        % x‑ and y‑label font
			tick label style={font=\small},   % every tick‑label font
		]

		\nextgroupplot[
			ylabel=Mbps,
			hide x axis,
			major tick length=0.2cm,
			axis line style=thick,
			tick align=outside,
			tick pos=left,
			axis y line*=left,
			ytick style={color=black, thick},
			y axis line style={->, >=latex},
			xmin=-4,xmax=292,
			xtick={0, 48, 96, 144, 192, 240, 288},
			xticklabels={00:00, 04:00, 08:00, 12:00, 16:00, 20:00, 24:00},
			ytick={2, 8}]
		\addplot [mark=no, Africa, thick]table[col sep=comma, x=index, y=data_in] {figures/three_profiles/data/250_traffic_profile.csv};

		\nextgroupplot[ylabel=Gbps,
			hide x axis,
			major tick length=0.2cm,
			axis line style=thick,
			tick align=outside,
			tick pos=left,
			axis y line*=left,
			ytick style={color=black, thick},
			y axis line style={->, >=latex},
			xmin=-4,xmax=292,
			xtick={0, 48, 96, 144, 192, 240, 288},
			xticklabels={00:00, 04:00, 08:00, 12:00, 16:00, 20:00, 24:00},
			ytick={4, 9}]
		\addplot [mark=no, Australia, thick]table[col sep=comma, x=index, y=data_in] {figures/three_profiles/data/34_traffic_profile.csv};

		\nextgroupplot[ylabel=Tbps,
			major tick length=0.2cm,
			axis line style=thick,
			tick align=outside,
			tick pos=left,
			axis y line*=left,
			ytick style={color=black, thick},
			xtick style={color=black, thick},
			axis x line*=bottom,
			xlabel={Local Time},
			y axis line style={->, >=latex},
			xmin=-4,xmax=292,
			xtick={0, 48, 96, 144, 192, 240, 288},
			xticklabels={00:00, 04:00, 08:00, 12:00, 16:00, 20:00, 24:00},
			ytick={1, 2}]
		\addplot [mark=no, Europe, thick]table[col sep=comma, x=index, y=data_in] {figures/three_profiles/data/297_traffic_profile.csv};

	\end{groupplot}
\end{tikzpicture}
		\caption{Same-day, local-time traffic for three IXPs, illustrating how diurnal patterns emerge as average volume increases.}\label{fig:three_profiles}
	\end{center}
\end{figure}

\section{Usage Patterns}\label{sec:usage-patterns}

Beyond volumetric trends, our dataset captures daily and weekly traffic rhythms that reflect synchronized Internet user behaviors.
This section examines the ``shape'' of traffic profiles in our dataset to understand underlying drivers of network demand and usage patterns.

We first investigate how discernible traffic patterns emerge.
We demonstrate that, as traffic volume passing through an IXP increases, observed traffic patterns shift from unstructured fluctuations to structured diurnal cycles~(\S\ref{subsec:from-noise-to-diurnal-rhythms}).
We then characterize and compare these typical rhythms across different regions~(\S\ref{subsec:pattern-regional-view}).
Finally, we show how significant deviations from these traffic patterns indicate real-world events affecting network usage~(\S\ref{subsec:deviations-from-norms}).

\subsection{The emergence of diurnal patterns}\label{subsec:from-noise-to-diurnal-rhythms}

Analysis of IXP traffic at varying scales reveals a fundamental trend: the emergence of clear, repeating structures in usage patterns correlates strongly with the aggregate traffic volume.
At lower volumes, typically below 1~Gbps, daily traffic profiles often appear as relatively unstructured fluctuations without a consistently repeating shape (top plot, \cref{fig:three_profiles}).
As traffic volume increases and more individual user activities are aggregated, these random variations tend to average out, and distinct diurnal patterns reflecting collective human activity typically become apparent (middle plot).
Once average traffic levels exceed approximately 10--15~Gbps, these daily and weekly patterns generally become well-defined and predictable (bottom plot).

To quantitatively assess when these fluctuations give way to repeating structures, we turn to information theory.
We use Shannon entropy (a measure of randomness) to characterize daily traffic patterns.
In our context, lower entropy indicates structured traffic patterns and the existence of diurnal cycles, whereas higher entropy reflects erratic and irregular usage throughout the day.
More intuitively, lower entropy indicates a more structured traffic pattern, where usage is concentrated in distinct periods, characteristic of clear diurnal cycles. Conversely, higher entropy reflects an erratic distribution of traffic across the day, suggesting flatter usage profiles without pronounced daily peaks and troughs.

Plotting Shannon entropy against average traffic rate for each traffic profile confirms this trend~(\cref{fig:spectral_entropy}).
Entropy generally decreases as traffic volume increases.
This demonstrates that, as IXPs aggregate more diverse user activities, the resulting collective behavior smooths into lower-entropy diurnal patterns.
\cref{fig:spectral_entropy} also reveals notable outliers: a cluster of vantage points characterized by both high traffic volumes and persistently high entropy.
Manual investigation identifies several of these as major research and education (R\&E) IXPs (e.g.\ CERN IX, PacificWave, and VIX) as depicted with red markers.
To our knowledge, these are the only R\&E IXPs represented in our dataset.
This suggests that these profiles may capture significant volumes of machine-driven traffic, such as large-scale scientific data transfers or inter-campus backups.
Unlike typical human-driven traffic, such automated flows can result in continuous or irregular bursts of data at any time, leading to a flatter traffic distribution and thereby maintaining higher entropy despite substantial volumes.
This suggests that metrics like Shannon entropy could differentiate human-driven and machine-driven traffic in networks, potentially enabling passive detection of botnets or other large-scale malicious activity.

\begin{figure}[t!]
	\begin{center}
		\begin{tikzpicture}
	\begin{scope}[line join=round]
		\begin{axis}[
				tick align=outside,
				tick pos=left,
				axis line style=thick,
				axis x line*=bottom,
				x axis line style={->, >=latex},
				%y axis line style={->, >=latex},
				axis y line*=left,
				major tick length=0.2cm,
				xtick style={color=black, thick},
				ytick style={color=black, thick},
				xlabel near ticks,                % x‑label tight to the axis
				ylabel near ticks,                % y‑label tight to the axis
				label style={font=\small},        % x‑ and y‑label font
				tick label style={font=\small},   % every tick‑label font
				ylabel={Entropy},
				xlabel={Mean traffic rate},
				xticklabel style={/pgf/number format/.cd, fixed, fixed zerofill, precision=1},
				height=4.5cm,
				width=0.90\linewidth,
				ymin=0, ymax=1,
				xmode=log, % Set x-axis to log scale
				xmin=5e-3, % Log scale cannot start at 0
				xmax=1e5,
				xtick={1e-2,1e0,1e2,1e4},
				xticklabels={10 Mbps, 1 Gbps, 100 Gbps, 10 Tbps},
				legend style={at={(0.5,1.25)}, anchor=north, draw=none, font=\small, legend columns=3, column sep=0.15cm},
				legend cell align={left}
			]

			\addplot+[only marks, mark=x, mark size=2pt, draw=GlobeDark] table[col sep=comma, x=mean_traffic_rate, y=spectral_entropy] {figures/spectral_entropy/data/single_entropy.csv};
			\addlegendentry{Single point}
			\addplot+[only marks, mark=o, mark size=2pt, draw=GlobeDark] table[col sep=comma, x=mean_traffic_rate, y=spectral_entropy] {figures/spectral_entropy/data/aggregated_entropy.csv};
			\addlegendentry{Federated}
			\addplot+[only marks, mark=x, mark size=3pt, ultra thick, draw=Africa, fill=Africa] table[col sep=comma, x=mean_traffic_rate, y=spectral_entropy] {figures/spectral_entropy/data/outliers.csv};
			\addlegendentry{R\&E IXPs}
		\end{axis}
	\end{scope}
\end{tikzpicture}
		\caption{As average traffic increases, daily patterns become more regular (lower spectral entropy), with notable outliers.}\label{fig:spectral_entropy}
	\end{center}
\end{figure}

\begin{figure}[t!]
	\begin{center}
		\begin{tikzpicture}
    \begin{axis}[
        tick align=outside,
        tick pos=left,
        axis line style=thick,
        axis x line*=bottom,
        x axis line style={},
        y axis line style={},
        axis y line*=left,
        major tick length=0.2cm,
        xtick style={color=black, thick, font=\small},
        ytick style={color=black, thick, font=\small},
        ylabel={Correlation},
        xlabel={Proportion of total traffic},
        xlabel near ticks,
        ylabel near ticks,
        label style={font=\small},
        ymin=0.8, ymax=1,
        xmin=0.01, xmax=1,
        xmode=log,
        xtick={0.01, 0.1, 1},
        xticklabels={1\%, 10\%, 100\%},
        ytick={0.8, 0.9, 1},
        yticklabels={0.8, 0.9, 1},
        height=4.5cm,
        width=0.90\linewidth,
        clip=false,
        set layers=standard,
        legend style={at={(0.42,1.05)}, anchor=south, legend columns=4, draw=none, font=\small, legend image post style={line width=2pt}, legend cell align=left, fill=none},
        legend cell align=left,
    ]
    
    \addplot[color=Africa, thick, mark=*, mark size=1pt] table[col sep=comma,x=x,y=y] {figures/day_corr/data/africa_correlation.csv};
    \addlegendentry{Africa}
    
    \addplot[color=AsiaPacific, thick, mark=*, mark size=1pt] table[col sep=comma,x=x,y=y] {figures/day_corr/data/asia_pacific_correlation.csv};
    \addlegendentry{Asia-Pacific}
    
    \addplot[color=Australia, thick, mark=*, mark size=1pt] table[col sep=comma,x=x,y=y] {figures/day_corr/data/australia_correlation.csv};
    \addlegendentry{Australia}
    
    \addplot[color=Europe, thick, mark=*, mark size=1pt] table[col sep=comma,x=x,y=y] {figures/day_corr/data/europe_correlation.csv};
    \addlegendentry{Europe}
    
    \addplot[color=MiddleEast, thick, mark=*, mark size=1pt] table[col sep=comma,x=x,y=y] {figures/day_corr/data/middle_east_correlation.csv};
    \addlegendentry{Middle East}
    
    \addplot[color=NorthAmerica, thick, mark=*, mark size=1pt] table[col sep=comma,x=x,y=y] {figures/day_corr/data/north_america_correlation.csv};
    \addlegendentry{North America}
    
    \addplot[color=SouthAmerica, thick, mark=*, mark size=1pt] table[col sep=comma,x=x,y=y] {figures/day_corr/data/south_america_correlation.csv};
    \addlegendentry{South America}
    
    \end{axis}
\end{tikzpicture}
		\caption{Regional weekly traffic patterns stabilize rapidly, often requiring less than 2\% of observed traffic to strongly correlate with the overall regional pattern.}\label{fig:day_corr}
	\end{center}
\end{figure}

\subsection{Characterizing regional patterns}\label{subsec:pattern-regional-view}

Across our dataset, the majority of our 378 traffic feeds from 472 tracked IXPs consistently exhibit diurnal patterns at sufficiently high traffic volumes, reflecting the rhythmic nature of Internet usage.
This subsection examines how diurnal patterns vary across regions in our dataset.

To analyze traffic pattern shapes, we adjust each profile, including single and federated profiles that are clustered by timezone, to their local time.
These adjusted profiles are then aggregated to derive regional traffic patterns.

To determine how representative an IXP's traffic pattern is of its region, we conduct a detailed analysis---analogous to the growth trend stability analysis in~\S\ref{subsec:growth-regional-view}---assessing the consistency of temporal patterns across IXPs within each region.
Specifically, starting with the IXP contributing the least traffic in a region and progressively adding larger ones by volume, we calculate the Pearson correlation between the weekly pattern of the growing partial aggregate and the final pattern derived from all observed IXPs in that region.
\cref{fig:day_corr} shows the percentage of observed regional traffic required to achieve near-perfect correlation with the final, fully aggregated regional pattern.

\newpage
The results reveal a striking contrast to the stabilization of growth trends. Characteristic weekly patterns emerge and converge almost immediately: in most regions, a correlation coefficient of $R>0.95$ is achieved with less than 2\% of observed traffic---often with under 0.1\%.
Even $R>0.99$ typically requires sampling under 10\%.
This rapid stabilization demonstrates that the weekly ``shape'' of Internet usage is a strong, consistent signal within a region, captured by observing only a small fraction (approximately 10--15~Gbps) of its traffic in our dataset.

\cref{fig:region_weeks} presents the average traffic profile for each day of the week, computed per region and normalized by the regional average traffic.
These profiles reveal distinct regional rhythms and highlight consistent contrasts between weekdays and weekends.

\begin{figure}[p]
	\centering
\begin{subfigure}[b]{\linewidth}
	\centering
	\begin{tikzpicture}

	\begin{scope}[line join=round]
		\begin{axis}[
				tick align=outside,
				tick pos=left,
				xmin=1, xmax=288,
				xtick={25, 73, 121, 169, 217, 265},
				xtick style={color=black, thick},
				xticklabels={},
				ymin=0.3, ymax=1.7,
				axis x line* = bottom,
				axis y line* = left,
				y axis line style={->, >=latex},
				major tick length=0.2cm,
				ytick={0.5, 1, 1.5},
				ytick style={color=black, thick},
				xticklabel style={font=\small},
				yticklabel style={font=\small},
				scaled x ticks=false,
				width=0.90\linewidth,
				height=3.9cm,
				legend style={
					at={(0.5,1.1)},
					anchor=south,
					legend columns=4,
					draw=none,
					column sep=0.1cm, 
					row sep=0.05cm, 
					font=\small, 
					legend image post style={line width=2pt}, 
					legend cell align=left,
				},
				legend cell align=left,
			]

			\addplot [thin, Monday] table[x=time_index, y=normalized_traffic, col sep=comma] {figures/weekly/south_america/data/monday.csv};
			\addplot [thin, Wednesday] table[x=time_index, y=normalized_traffic, col sep=comma] {figures/weekly/south_america/data/wednesday.csv};
			\addplot [thin, Friday] table[x=time_index, y=normalized_traffic, col sep=comma] {figures/weekly/south_america/data/friday.csv};
			\addplot [thin, Sunday] table[x=time_index, y=normalized_traffic, col sep=comma] {figures/weekly/south_america/data/sunday.csv};		
			\addplot [thin, Tuesday] table[x=time_index, y=normalized_traffic, col sep=comma] {figures/weekly/south_america/data/tuesday.csv};
			\addplot [thin, Thursday] table[x=time_index, y=normalized_traffic, col sep=comma] {figures/weekly/south_america/data/thursday.csv};
			\addplot [thin, Saturday] table[x=time_index, y=normalized_traffic, col sep=comma] {figures/weekly/south_america/data/saturday.csv};

			\legend{Monday, Wednesday, Friday, Sunday, Tuesday, Thursday, Saturday}
		\end{axis}
		
		\node[anchor=north west, font=\normalsize] at ([xshift=1em]current axis.north west) {South America};
	\end{scope}
\end{tikzpicture}
\end{subfigure}
\begin{subfigure}[b]{\linewidth}
	\centering
	\begin{tikzpicture}

	\begin{scope}[line join=round]
		\begin{axis}[
				tick align=outside,
				tick pos=left,
				xmin=1, xmax=288,
				xtick={25, 73, 121, 169, 217, 265},
				xtick style={color=black, thick},
				xticklabels={},
				ymin=0.3, ymax=1.7,
				axis x line* = bottom,
				axis y line* = left,
				y axis line style={->, >=latex},
				major tick length=0.2cm,
				ytick={0.5, 1, 1.5},
				ytick style={color=black, thick},
				xticklabel style={font=\small},
				yticklabel style={font=\small},
				scaled x ticks=false,
				width=0.90\linewidth,
				height=3.9cm,
			]

			\addplot [thin, Monday] table[x=time_index, y=normalized_traffic, col sep=comma] {figures/weekly/middle_east/data/monday.csv};
			\addplot [thin, Tuesday] table[x=time_index, y=normalized_traffic, col sep=comma] {figures/weekly/middle_east/data/tuesday.csv};
			\addplot [thin, Wednesday] table[x=time_index, y=normalized_traffic, col sep=comma] {figures/weekly/middle_east/data/wednesday.csv};
			\addplot [thin, Thursday] table[x=time_index, y=normalized_traffic, col sep=comma] {figures/weekly/middle_east/data/thursday.csv};
			\addplot [thin, Friday] table[x=time_index, y=normalized_traffic, col sep=comma] {figures/weekly/middle_east/data/friday.csv};	
			\addplot [thin, Saturday] table[x=time_index, y=normalized_traffic, col sep=comma] {figures/weekly/middle_east/data/saturday.csv};
			\addplot [thin, Sunday] table[x=time_index, y=normalized_traffic, col sep=comma] {figures/weekly/middle_east/data/sunday.csv};	
		\end{axis}
		
		\node[anchor=north west, font=\normalsize] at ([xshift=1em]current axis.north west) {Middle East};
	\end{scope}
\end{tikzpicture}
\end{subfigure}
\begin{subfigure}[b]{\linewidth}
	\centering
	\begin{tikzpicture}

	\begin{scope}[line join=round]
		\begin{axis}[
				tick align=outside,
				tick pos=left,
				xmin=1, xmax=288,
				xtick={25, 73, 121, 169, 217, 265},
				xtick style={color=black, thick},
				xticklabels={},
				ymin=0.3, ymax=1.7,
				axis x line* = bottom,
				axis y line* = left,
				y axis line style={->, >=latex},
				major tick length=0.2cm,
				ytick={0.5, 1, 1.5},
				ytick style={color=black, thick},
				xticklabel style={font=\small},
				yticklabel style={font=\small},
				scaled x ticks=false,
				width=0.90\linewidth,
				height=3.9cm,
			]

			\addplot [thin, Monday] table[x=time_index, y=normalized_traffic, col sep=comma] {figures/weekly/asia_pacific/data/monday.csv};
			\addplot [thin, Tuesday] table[x=time_index, y=normalized_traffic, col sep=comma] {figures/weekly/asia_pacific/data/tuesday.csv};
			\addplot [thin, Wednesday] table[x=time_index, y=normalized_traffic, col sep=comma] {figures/weekly/asia_pacific/data/wednesday.csv};
			\addplot [thin, Thursday] table[x=time_index, y=normalized_traffic, col sep=comma] {figures/weekly/asia_pacific/data/thursday.csv};
			\addplot [thin, Friday] table[x=time_index, y=normalized_traffic, col sep=comma] {figures/weekly/asia_pacific/data/friday.csv};	
			\addplot [thin, Saturday] table[x=time_index, y=normalized_traffic, col sep=comma] {figures/weekly/asia_pacific/data/saturday.csv};
			\addplot [thin, Sunday] table[x=time_index, y=normalized_traffic, col sep=comma] {figures/weekly/asia_pacific/data/sunday.csv};	
		\end{axis}
		
		\node[anchor=north west, font=\normalsize] at ([xshift=1em]current axis.north west) {Asia-Pacific};
	\end{scope}
\end{tikzpicture}
\end{subfigure}
\begin{subfigure}[b]{\linewidth}
	\centering
	\begin{tikzpicture}

	\begin{scope}[line join=round]
		\begin{axis}[
				tick align=outside,
				tick pos=left,
				xmin=1, xmax=288,
				xtick={25, 73, 121, 169, 217, 265},
				xtick style={color=black, thick},
				xticklabels={},
				ymin=0.3, ymax=1.7,
				axis x line* = bottom,
				axis y line* = left,
				y axis line style={->, >=latex},
				major tick length=0.2cm,
				ytick={0.5, 1, 1.5},
				ytick style={color=black, thick},
				xticklabel style={font=\small},
				yticklabel style={font=\small},
				scaled x ticks=false,
				width=0.90\linewidth,
				height=3.9cm,
			]

			\addplot [thin, Monday] table[x=time_index, y=normalized_traffic, col sep=comma] {figures/weekly/australia/data/monday.csv};
			\addplot [thin, Tuesday] table[x=time_index, y=normalized_traffic, col sep=comma] {figures/weekly/australia/data/tuesday.csv};
			\addplot [thin, Wednesday] table[x=time_index, y=normalized_traffic, col sep=comma] {figures/weekly/australia/data/wednesday.csv};
			\addplot [thin, Thursday] table[x=time_index, y=normalized_traffic, col sep=comma] {figures/weekly/australia/data/thursday.csv};
			\addplot [thin, Friday] table[x=time_index, y=normalized_traffic, col sep=comma] {figures/weekly/australia/data/friday.csv};	
			\addplot [thin, Saturday] table[x=time_index, y=normalized_traffic, col sep=comma] {figures/weekly/australia/data/saturday.csv};
			\addplot [thin, Sunday] table[x=time_index, y=normalized_traffic, col sep=comma] {figures/weekly/australia/data/sunday.csv};	
		\end{axis}
		
		\node[anchor=north west, font=\normalsize] at ([xshift=1em]current axis.north west) {Australia};
	\end{scope}
\end{tikzpicture}
\end{subfigure}
\begin{subfigure}[b]{\linewidth}
	\centering
	\begin{tikzpicture}

	\begin{scope}[line join=round]
		\begin{axis}[
				tick align=outside,
				tick pos=left,
				xmin=1, xmax=288,
				xtick={25, 73, 121, 169, 217, 265},
				xtick style={color=black, thick},
				xticklabels={},
				ymin=0.3, ymax=1.7,
				axis x line* = bottom,
				axis y line* = left,
				y axis line style={->, >=latex},
				major tick length=0.2cm,
				ytick={0.5, 1, 1.5},
				ytick style={color=black, thick},
				xticklabel style={font=\small},
				yticklabel style={font=\small},
				scaled x ticks=false,
				width=0.90\linewidth,
				height=3.9cm,
			]

			\addplot [thin, Monday] table[x=time_index, y=normalized_traffic, col sep=comma] {figures/weekly/europe/data/monday.csv};
			\addplot [thin, Tuesday] table[x=time_index, y=normalized_traffic, col sep=comma] {figures/weekly/europe/data/tuesday.csv};
			\addplot [thin, Wednesday] table[x=time_index, y=normalized_traffic, col sep=comma] {figures/weekly/europe/data/wednesday.csv};
			\addplot [thin, Thursday] table[x=time_index, y=normalized_traffic, col sep=comma] {figures/weekly/europe/data/thursday.csv};
			\addplot [thin, Friday] table[x=time_index, y=normalized_traffic, col sep=comma] {figures/weekly/europe/data/friday.csv};	
			\addplot [thin, Saturday] table[x=time_index, y=normalized_traffic, col sep=comma] {figures/weekly/europe/data/saturday.csv};
			\addplot [thin, Sunday] table[x=time_index, y=normalized_traffic, col sep=comma] {figures/weekly/europe/data/sunday.csv};	
		\end{axis}
		
		\node[anchor=north west, font=\normalsize] at ([xshift=1em]current axis.north west) {Europe};
	\end{scope}
\end{tikzpicture}
\end{subfigure}
\begin{subfigure}[b]{\linewidth}
	\centering
	\begin{tikzpicture}

	\begin{scope}[line join=round]
		\begin{axis}[
				tick align=outside,
				tick pos=left,
				xmin=1, xmax=288,
				xtick={25, 73, 121, 169, 217, 265},
				xtick style={color=black, thick},
				xticklabels={},
				ymin=0.3, ymax=1.7,
				axis x line* = bottom,
				axis y line* = left,
				y axis line style={->, >=latex},
				major tick length=0.2cm,
				ytick={0.5, 1, 1.5},
				ytick style={color=black, thick},
				xticklabel style={font=\small},
				yticklabel style={font=\small},
				scaled x ticks=false,
				width=0.90\linewidth,
				height=3.9cm,
			]

			\addplot [thin, Monday] table[x=time_index, y=normalized_traffic, col sep=comma] {figures/weekly/north_america/data/monday.csv};
			\addplot [thin, Tuesday] table[x=time_index, y=normalized_traffic, col sep=comma] {figures/weekly/north_america/data/tuesday.csv};
			\addplot [thin, Wednesday] table[x=time_index, y=normalized_traffic, col sep=comma] {figures/weekly/north_america/data/wednesday.csv};
			\addplot [thin, Thursday] table[x=time_index, y=normalized_traffic, col sep=comma] {figures/weekly/north_america/data/thursday.csv};
			\addplot [thin, Friday] table[x=time_index, y=normalized_traffic, col sep=comma] {figures/weekly/north_america/data/friday.csv};	
			\addplot [thin, Saturday] table[x=time_index, y=normalized_traffic, col sep=comma] {figures/weekly/north_america/data/saturday.csv};
			\addplot [thin, Sunday] table[x=time_index, y=normalized_traffic, col sep=comma] {figures/weekly/north_america/data/sunday.csv};	
		\end{axis}
		
		\node[anchor=north west, font=\normalsize] at ([xshift=1em]current axis.north west) {North America};
	\end{scope}
\end{tikzpicture}
\end{subfigure}
\begin{subfigure}[b]{\linewidth}
	\centering
	\begin{tikzpicture}

	\begin{scope}[line join=round]
		\begin{axis}[
				tick align=outside,
				tick pos=left,
				xmin=1, xmax=288,
				xtick={25, 73, 121, 169, 217, 265},
				xticklabels={02:00, 06:00, 10:00, 14:00, 18:00, 22:00},
				xtick style={color=black, thick},
				ymin=0.3, ymax=1.7,
				axis x line* = bottom,
				axis y line* = left,
				y axis line style={->, >=latex},
				major tick length=0.2cm,
				ytick={0.5, 1, 1.5},
				ytick style={color=black, thick},
				xticklabel style={font=\small},
				yticklabel style={font=\small},
				scaled x ticks=false,
				width=0.90\linewidth,
				height=3.9cm,
			]

			\addplot [thin, Monday] table[x=time_index, y=normalized_traffic, col sep=comma] {figures/weekly/africa/data/monday.csv};
			\addplot [thin, Wednesday] table[x=time_index, y=normalized_traffic, col sep=comma] {figures/weekly/africa/data/wednesday.csv};
			\addplot [thin, Friday] table[x=time_index, y=normalized_traffic, col sep=comma] {figures/weekly/africa/data/friday.csv};
			\addplot [thin, Sunday] table[x=time_index, y=normalized_traffic, col sep=comma] {figures/weekly/africa/data/sunday.csv};		
			\addplot [thin, Tuesday] table[x=time_index, y=normalized_traffic, col sep=comma] {figures/weekly/africa/data/tuesday.csv};
			\addplot [thin, Thursday] table[x=time_index, y=normalized_traffic, col sep=comma] {figures/weekly/africa/data/thursday.csv};
			\addplot [thin, Saturday] table[x=time_index, y=normalized_traffic, col sep=comma] {figures/weekly/africa/data/saturday.csv};
		\end{axis}
		
		\node[anchor=north west, font=\normalsize] at ([xshift=1em]current axis.north west) {Africa};
	\end{scope}
\end{tikzpicture}
\end{subfigure}
\caption{Distinct weekly traffic patterns across regions (normalized by each region's mean), reflecting cultural schedules and usage behaviors.}\label{fig:region_weeks}
\end{figure}

\vspace{0.5em}\noindent\emph{South America---the boom-and-bust continent.}
On weekdays, the traffic surges with a peak-to-trough ratio of 4.7$\times$---a full 40\% higher than any other region.
Yet by Saturday, the peak diminishes by 8.8\%, marking the steepest weekend drop observed.
The trough is equally pronounced, bottoming out at 0.34$\times$ the regional mean at 05:10.
Marked by explosive evenings and silent nights, South American traffic is emblematic of a residential-heavy region.

\vspace{0.5em}\noindent\emph{Middle East---where night is the prime time.}
Cultural and religious customs shift the weekend to Friday-Saturday---and the Internet follows.
The weekday peak hits at 18:15, but on the regional weekend, it shifts a remarkable 4~hours and 35~minutes later, to 22:50.
Despite this shift, the amplitude remains below 1.9$\times$---the flattest profile observed in our dataset.
More striking still, the \textit{late-night share} (23:00-02:00) rises 17\% above the daily mean, the highest worldwide.
When we add the latest trough (07:35), a pattern emerges: users stay up---and wake up---later than anywhere else.

\vspace{0.5em}\noindent\emph{Asia-Pacific---always on, scarcely off.}
In Asia-Pacific, the weekend barely registers: peak traffic drops just 3.7\%, and the peak time shifts by just 5~minutes.
Amplitudes stay high (weekday 3.47$\times$, weekend 3.34$\times$), but the trough never dips below 0.42$\times$.
Traffic barely dips after midnight, with troughs arriving at 04:45.

\vspace{0.5em}\noindent\emph{Australia---early birds with sturdy weekends.}
Down Under, the evening peak arrives earliest at 20:25, and---uniquely---rises slightly on weekends (0.28\%).
Weekday amplitude reaches 3.47$\times$, dipping only slightly to 3.34$\times$ on Saturday and Sunday.
During the 08:00-12:00 window, weekend traffic averages approximately 10\% below weekday levels---twice the mid-morning dip in Africa (5\%) and exceeding other regions (two of which show increases).
In short, while much of the world eases into its Saturday scroll, users in Australia are emphatically offline.

\vspace{0.5em}\noindent\emph{Europe vs. North America---same clock, different motors.}
Both regions hit their trough at 04:05, but that's where the symmetry ends.
Europe exhibits a sharper weekday amplitude (3.15$\times$) and a modest weekend dip (1.6\%), while North America flattens out with a lower amplitude (2.19$\times$) and virtually no weekend change (+0.2\%).

\vspace{0.5em}\noindent\emph{Africa---weekdays wired, weekends rising.}
Africa's peaks barely move (-0.05\% weekend change), yet weekday amplitude hits a solid 3.26$\times$.
The trough (04:14, 0.41$\times$) is deeper than Europe's but less extreme than South America's.

\vspace{0.5em}
\noindent These regional traffic patterns demonstrate that while the fundamental diurnal rhythm of Internet usage is universal, its specific characteristics vary significantly across regions. The consistency of these patterns within regions, combined with their distinctiveness across regions, underscores the richness of IXP traffic data.

\subsection{Detecting outliers in diurnal patterns}\label{subsec:deviations-from-norms}

The inherent stability of daily and weekly usage patterns, established in the preceding analyses, provides a robust baseline against which significant deviations can be identified.
Such deviations can signal anomalous network events driven by impactful real-world occurrences.
Building on this observed pattern regularity, we identify anomalous days by detecting substantial variations in their daily traffic shape compared to a region's typical profile for that day across other weeks.
Once an anomalous day is flagged, its intra-day traffic curve is examined against the regional average to pinpoint specific event timings and quantify their impact.
For this paper, we validate and select outliers manually; nonetheless, we observe hundreds of such candidates across regions.
Work is underway on an automated, scalable pipeline to discover, validate, and characterize these events systematically.

Our dataset reveals numerous instances where aggregated IXP traffic patterns clearly reflect the impact of major societal events.
We present a manually curated subset in~\cref{fig:outliers}, contrasting the traffic profile on the event day with typical patterns.
For instance, in Africa, traffic on July 26, 2024, deviated significantly from the regional norm, coinciding with the Olympic Games opening ceremony, with observed traffic peaking at twice the typical Friday evening level during ceremony hours.
A similar impact was observed in South America on August 7, 2024, during the Olympics Women's football semi-final (Spain vs. Brazil), a popular derby, where traffic during the match hours rose to nearly twice the usual volume for that period.
The A-League Men's Grand Final in Australia (May 24, 2024) also produced a noticeable, albeit more moderate, traffic increase of 20-25\% during match time (dashed line).

\begin{figure}[t!]
	\centering
\begin{subfigure}[b]{\linewidth}
	\centering
	\begin{tikzpicture}

	\begin{scope}[line join=round]
		\begin{axis}[
				tick align=outside,
				tick pos=left,
				xmin=1, xmax=864,
				xtick={1, 289, 577},
				xtick style={color=black, thick},
				xticklabels={},
				ymin=0, ymax=15,
				axis x line* = bottom,
				axis y line* = left,
				y axis line style={->, >=latex},
				major tick length=0.2cm,
				ytick={0, 3, 6, 9, 12},
				ytick style={color=black, thick},
				xticklabel style={font=\small},
				yticklabel style={font=\small},
				ylabel={Traffic (Tbps)},
				scaled x ticks=false,
				width=0.90\linewidth,
				height=3.5cm,
				xmajorgrids=true,
				grid style={dashed, black},
				legend style={at={(0.03,0.75)}, anchor= south west, draw=none, fill=white, legend columns=1, row sep=0.05cm, font=\small, legend image post style={line width=2pt}},
				legend cell align=left		
			]

			\addplot [thin, Africa] table[x=sequence, y=total_traffic_tb, col sep=comma] {figures/outliers/africa/data/olympics.csv};
			\addlegendentry{Olympics Ceremony}
			
		\end{axis}
		
		\node[anchor=north east, font=\normalsize] at (current axis.north east) {Africa};
	\end{scope}
\end{tikzpicture}
\end{subfigure}
\begin{subfigure}[b]{\linewidth}
	\centering
	\begin{tikzpicture}

	\begin{scope}[line join=round]
		\begin{axis}[
				tick align=outside,
				tick pos=left,
				xmin=1, xmax=864,
				xtick={1, 289, 577},
				xtick style={color=black, thick},
				xticklabels={},
				ymin=0, ymax=5,
				axis x line* = bottom,
				axis y line* = left,
				y axis line style={->, >=latex},
				major tick length=0.2cm,
				ytick={0, 1, 2, 3, 4},
				ytick style={color=black, thick},
				xticklabel style={font=\small},
				yticklabel style={font=\small},
				ylabel={Traffic (Tbps)},
				scaled x ticks=false,
				width=0.90\linewidth,
				height=3.5cm,
				xmajorgrids=true,
				grid style={dashed, black},
				legend style={at={(0.03,0.75)}, anchor= south west, draw=none, fill=white, legend columns=1, row sep=0.05cm, font=\small, legend image post style={line width=2pt}},
				legend cell align=left		
			]

			\addplot [thin, Australia] table[x=sequence, y=total_traffic_tb, col sep=comma] {figures/outliers/australia/data/fortnite.csv};
			\addlegendentry{Fortnite}

			\addplot [thin, Australia, dashed] table[x=sequence, y=total_traffic_tb, col sep=comma] {figures/outliers/australia/data/football.csv};
			\addlegendentry{A-League Final}
			
		\end{axis}
		
		\node[anchor=north east, font=\normalsize] at (current axis.north east) {Australia};
	\end{scope}
\end{tikzpicture}
\end{subfigure}
\begin{subfigure}[b]{\linewidth}
	\centering
	\begin{tikzpicture}

	\begin{scope}[line join=round]
		\begin{axis}[
				tick align=outside,
				tick pos=left,
				xmin=1, xmax=865,
				xtick={1, 289, 577, 865},
				xtick style={color=black, thick},
				x tick label as interval, 
				xticklabels={Pre-event day, Event day, Post-event day},
				ymin=0, ymax=90,
				axis x line* = bottom,
				axis y line* = left,
				y axis line style={->, >=latex},
				major tick length=0.2cm,
				ytick={0, 20, 40, 60, 80},
				ytick style={color=black, thick},
				xticklabel style={font=\small},
				yticklabel style={font=\small},
				ylabel={Traffic (Tbps)},
				scaled x ticks=false,
				width=0.90\linewidth,
				height=3.5cm,
				legend style={at={(0.03,0.75)}, anchor= south west, draw=none, fill=white, legend columns=1, row sep=0.05cm, font=\small, legend image post style={line width=2pt}},
				legend cell align=left	
			]

			\addplot [thin, SouthAmerica] table[x=sequence, y=total_traffic_tb, col sep=comma] {figures/outliers/south_america/data/football.csv};
			\addlegendentry{Olympics Derby}

			\addplot [thin, SouthAmerica, dashed] table[x=sequence, y=total_traffic_tb, col sep=comma] {figures/outliers/south_america/data/reality.csv};
			\addlegendentry{Big Brother Brasil}

			% --- draw ONLY the two desired vertical grid lines
			\draw[dashed, black] (axis cs:289,0) -- (axis cs:289,90);
			\draw[dashed, black] (axis cs:577,0) -- (axis cs:577,90);

		\end{axis}
		
		\node[anchor=north east, font=\normalsize] at (current axis.north east) {South America};
	\end{scope}
\end{tikzpicture}
\end{subfigure}
	\caption{Sensitivity of regional IXP traffic patterns to major real-world events, showcasing distinct deviations from typical daily profiles.}\label{fig:outliers}
\end{figure}

Gaming events also leave a clear imprint.
In Australia, a major Fortnite update on November 30, 2024, led to sustained traffic elevation on December 1, 2024, with peak download and gaming hours showing traffic levels over twice the typical Sunday afternoon level.
This event also had a significant impact in Europe on December 1, 2024, with observed traffic during peak gaming periods approximately 10\% above typical Sunday levels.
The re-release of Fortnite's ``OG'' mode on December 6, 2024, had a further widespread impact, visibly elevating traffic in the Middle East (approximately 25\% above the typical baseline during evening gaming) and similarly affecting North America (around 20\% increase during peak engagement hours).
E-commerce events, such as Singles' Day on November 13, 2024, in Asia-Pacific, resulted in sustained traffic elevation throughout the day, approximately 30-40\% above the regional average for a comparable weekday.
Reality television, such as the Big Brother Brasil finale on April 4, 2024, also drove a significant traffic increase in our dataset.

These diverse examples demonstrate that aggregated IXP traffic patterns are highly sensitive and responsive to a wide range of real-world events that drive collective online activity.
The ability to detect and quantify these deviations validates the richness of IXP traffic data beyond simple volume metrics.
Such pattern deviation analysis not only provides insights into the network-level impact of different societal phenomena but also highlights the potential for using this approach in near-real-time monitoring to identify widespread service-demand shifts or other significant network anomalies.
Understanding the shape of traffic, therefore, offers a complementary and often more nuanced view of Internet dynamics compared to volumetric analysis alone.

\section{Discussion}\label{sec:discussion}

Having introduced our dataset and examined its characteristics across regions, we now discuss its implications: what applications and future research it enables, and its inherent limitations---what it \emph{cannot} do.

\paragraph*{Scope and limits of what we measure.}
At its peak, we observe more than 300~Tbps of IXP traffic in aggregate. However, we do not know what fraction of all Internet traffic this represents, nor how that fraction varies by region. The reported values capture only traffic on the \emph{public} IXP fabric; private network interconnections (PNIs) are outside our measurement scope and may constitute a substantial share in some exchanges.

As argued in \S\ref{ssec:representativeness-via-port-capacity}, however, capacity metadata indicates our visibility covers the vast majority of \emph{public} IXP capacity for each region---providing strong support for representativeness of public-fabric dynamics. Prior IXP studies reported tens of Tbps at major exchanges~\cite{10.1145/2342356.2342393,10.1145/2382016.2382018} and found IXP traffic to be diverse and broadly representative of Internet exchange activity~\cite{chatzisBenefitsUsingLarge2013,bottgerOpenConnectEverywhere2018}; our contribution is to provide this view at broad, longitudinal scale using public data instead of one-off collaborations.

\paragraph*{What our dataset enables.}
First, it offers a transparent, application-agnostic baseline for monitoring Internet growth. Our measured annual growth (23.4\% in 2023; 16.9\% in 2024) closely aligns with independent reports such as Cloudflare Radar (25\%, 17.2\%)~\cite{cloudflare_year_in_review_2023,cloudflare_year_in_review_2024} and with longer-horizon industry forecasts (e.g., Cisco's 20--30\% CAGR)~\cite{cisco_air_2018_2023}; Sandvine's reports on video's growth (24\% in 2023) are consistent with our aggregate increase~\cite{marwaha_gipr_2024,sandvine_gipr_pr_2023}. While those sources rely on proprietary telemetry, our public-IXP vantage provides a complementary, verifiable baseline. This complementarity is evident during the Paris 2024 Olympics opening ceremony: we observe traffic spikes at African IXPs, whereas Cloudflare reports a pronounced drop on its network~\cite{tomeParis2024Olympics2024}---consistent with live-streaming load appearing at IXPs while other Cloudflare-served content temporarily declines---illustrating how different blind men perceive different parts of the elephant.

Second, it enables \emph{regional and seasonal analyses} that are often absent from aggregate reports: a high-level view of regional and country-level Internet traffic growth and patterns can inform policy-makers and researchers about socio-economic and technological trends in a given geography. Furthermore, these reports usually lack the traffic shape and patterns that can inform network management, more sustainable infrastructure solutions, and investment and provisioning---enabling robust regional ``shape'' baselines useful for capacity planning, energy-aware operations, and benchmarking. Even where overall Internet visibility is limited, the IXP vantage can illuminate underrepresented regions; for Africa in particular, the prevalence of IXPs makes them a useful proxy for overall traffic~\cite{10.1145/3646547.3689679}.

Third, understanding usage patterns, or what normal traffic looks like at a vantage point, is especially important to detect \emph{outliers}. Our future work will cover detecting, documenting, and characterizing the outliers we see worldwide. Such a dataset enables research on the reliability of the Internet, the prevalence of security or failure challenges, and informs policy-makers in this age where the Internet has become a human right and a non-negotiable piece of modern life.

\paragraph*{What the dataset cannot do.}
Our vantage cannot \emph{(i)}~quantify PNI traffic or infer total interconnection volume; \emph{(ii)}~reveal application mix or content/provider shares; \emph{(iii)}~attribute causes from traffic alone (corroboration is needed); \emph{(iv)}~provide per-AS, per-flow, or path-level insights; or \emph{(v)}~guarantee uniform timeliness (a small number of feeds exhibit batching/lag, \S\ref{sec:geographic-coverage-and-inherent-biases}). Federated feeds can also blur regional assignment in a few cases (e.g., Equinix aggregates), and regional coverage remains lower in some markets (notably Asia-Pacific and North America; \S\ref{sec:geographic-coverage-and-inherent-biases}). Accordingly, we interpret capacity coverage as an \emph{upper bound} on provisioned public-fabric capacity, and avoid extrapolating to the entirety of Internet traffic.

\paragraph*{Ethics and responsible use.}
We use only publicly available, aggregate IXP statistics; we collect no packet-, flow-, or AS-level data and perform no de-anonymization. Scrapers honor robots.txt and Terms of Service, apply conservative polling/backoff and caching to minimize load, and never bypass authentication. We respect operator requests to update or withdraw feeds, log provenance for reproducibility, and release only aggregate, non-sensitive artifacts. Analyses avoid attributing behavior to individual networks without independent corroboration, and we invite community feedback to correct or retire feeds if circumstances change.

\paragraph*{Future directions.}
In addition to the automated event-detection pipeline, we see opportunities to: \emph{(i)}~improve coverage in opaque markets; \emph{(ii)}~model PNI dynamics indirectly (e.g., combining PeeringDB port churn, BGP announcements, and capacity utilization to bound PNI growth); \emph{(iii)}~integrate energy/CO\textsubscript{2} models with diurnal profiles for sustainability planning; and \emph{(iv)}~study resilience by correlating traffic anomalies with routing incidents and outages~\cite{10.1007/978-3-031-56252-5_10,bertholdo2021forecasting}. These directions build on the strengths of a public, longitudinal, broad IXP vantage while respecting its scope limits.

\paragraph*{The network evolves in lockstep with demand.} One of the remarkable things we observed (but do not present in detail due to space constraints) is how closely the port capacity of these IXPs grows with the growth in overall traffic.
Our investigations suggest that if this trend is consistent, port capacity could be used as a rough proxy for traffic growth even in IXPs that do not make their traffic statistics publicly available. We provide a deeper analysis of this trend in the supplementary materials~(\S\ref{sec:supplementary}).

\section{Conclusion}\label{sec:conclusion}

We present a longitudinal study of traffic volumes exchanged at IXPs around the world from January 2023 to December 2024. We demonstrate that IXP data, like the other blind men, has its own blind spots and biases, yet it forms a strong complement to our current understanding of Internet traffic. Not only is IXP data broadly distributed, available, and verifiable, but it is also fine-grained enough to capture regional usage patterns and atypical network usage in response to major events. Using the \emph{daily mean} as our canonical metric, we also observed aggregate traffic rise from 138~Tbps to 200~Tbps over 2023--2024 (+49.2\%; CAGR $\approx$\,24.5\%), which is in line with many other estimates of traffic growth on the Internet.

Crucially, this work establishes the validated baseline---broad geographic coverage, stable regional patterns, and well-characterized diurnal and seasonal behavior---that is a prerequisite for systematic anomaly detection. With this foundation in place, our ongoing work focuses on building automated pipelines for detecting and attributing global-scale network events, outages, and disruptions in near real-time.

\paragraph*{Availability.} The dataset is available at \url{https://github.com/nsg-ethz/ixp-traffic-dataset}.

%%
%% Bibliography
%%

%% Please use bibtex,

\clearpage
\bibliography{paper}

\begin{thebibliography}{10}

\bibitem{10.1145/2342356.2342393}
Bernhard Ager, Nikolaos Chatzis, Anja Feldmann, Nadi Sarrar, Steve Uhlig, and
  Walter Willinger.
\newblock Anatomy of a large european {{IXP}}.
\newblock In {\em Proceedings of the {{ACM SIGCOMM}} 2012 Conference on
  Applications, Technologies, Architectures, and Protocols for Computer
  Communication}, Sigcomm '12, pages 163--174, New York, NY, USA, 2012.
  Association for Computing Machinery.
\newblock \href {https://doi.org/10.1145/2342356.2342393}
  {\path{doi:10.1145/2342356.2342393}}.

\bibitem{cloudflare_year_in_review_2023}
David Belson.
\newblock Cloudflare 2023 {{Year}} in {{Review}}, December 2023.
\newblock URL: \url{https://blog.cloudflare.com/radar-2023-year-in-review/}.

\bibitem{cloudflare_year_in_review_2024}
David Belson.
\newblock Cloudflare 2024 {{Year}} in {{Review}}, December 2024.
\newblock URL: \url{https://blog.cloudflare.com/radar-2024-year-in-review/}.

\bibitem{bertholdo2021forecasting}
Leandro~Marcio Bertholdo, Jo{\~a}o~M Ceron, Lisandro~Zambenedetti Granville,
  and Roland {van Rijswijk-Deij}.
\newblock Forecasting the impact of {{IXP}} outages using anycast.
\newblock In {\em {{TMA}}}, 2021.

\bibitem{bottger2018shaping}
Timm B{\"o}ttger, Gianni Antichi, Eder Le{\~a}o~Fernandes, Roberto {di Lallo},
  Marc De~Bruy{\`e}re, Steve Uhlig, Gareth Tyson, and Ignacio Castro.
\newblock Shaping the internet: 10 years of {{IXP}} growth.
\newblock {\em arXiv (Cornell University)}, October 2018.
\newblock \href {https://doi.org/10.48550/arxiv.1810.10963}
  {\path{doi:10.48550/arxiv.1810.10963}}.

\bibitem{bottgerOpenConnectEverywhere2018}
Timm B{\"o}ttger, Felix Cuadrado, Gareth Tyson, Ignacio Castro, and Steve
  Uhlig.
\newblock Open {{Connect Everywhere}}: {{A Glimpse}} at the {{Internet
  Ecosystem}} through the {{Lens}} of the {{Netflix CDN}}.
\newblock {\em SIGCOMM Comput. Commun. Rev.}, 48(1):28--34, April 2018.
\newblock URL: \url{https://dl.acm.org/doi/10.1145/3211852.3211857}, \href
  {https://doi.org/10.1145/3211852.3211857}
  {\path{doi:10.1145/3211852.3211857}}.

\bibitem{10.1145/3276799.3276801}
Timm B{\"o}ttger, Felix Cuadrado, and Steve Uhlig.
\newblock Looking for hypergiants in {{peeringDB}}.
\newblock {\em SIGCOMM Comput. Commun. Rev.}, 48(3):13--19, September 2018.
\newblock \href {https://doi.org/10.1145/3276799.3276801}
  {\path{doi:10.1145/3276799.3276801}}.

\bibitem{bottgerHowInternetReacted2020}
Timm B{\"o}ttger, Ghida Ibrahim, and Ben Vallis.
\newblock How the {{Internet}} reacted to {{Covid-19}}: {{A}} perspective from
  {{Facebook}}'s {{Edge Network}}.
\newblock In {\em Proceedings of the {{ACM Internet Measurement Conference}}},
  {{IMC}} '20, pages 34--41, New York, NY, USA, October 2020. Association for
  Computing Machinery.
\newblock URL: \url{https://dl.acm.org/doi/10.1145/3419394.3423621}, \href
  {https://doi.org/10.1145/3419394.3423621}
  {\path{doi:10.1145/3419394.3423621}}.

\bibitem{brown2024rcs}
Lloyd Brown, Albert~Gran Alcoz, Frank Cangialosi, Akshay Narayan, Mohammad
  Alizadeh, Hari Balakrishnan, Eric Friedman, Ethan {Katz-Bassett}, Arvind
  Krishnamurthy, Michael Schapira, and Scott Shenker.
\newblock Principles for {{Internet Congestion Management}}.
\newblock In {\em Proceedings of the {{ACM SIGCOMM}} 2024 {{Conference}}},
  {{ACM SIGCOMM}} '24, pages 166--180, New York, NY, USA, August 2024.
  Association for Computing Machinery.
\newblock URL: \url{https://dl.acm.org/doi/10.1145/3651890.3672247}, \href
  {https://doi.org/10.1145/3651890.3672247}
  {\path{doi:10.1145/3651890.3672247}}.

\bibitem{caida_peeringdb_2023_2024}
{CAIDA UCSD}.
\newblock {{CAIDA UCSD PeeringDB}} dataset, 2023-01-01 -- 2024-12-31, 2017.
\newblock URL: \url{https://www.caida.org/catalog/datasets/peeringdb/}.

\bibitem{CANDELA2020107495}
Massimo Candela, Valerio Luconi, and Alessio Vecchio.
\newblock Impact of the {{COVID-19}} pandemic on the {{Internet}} latency:
  {{A}} large-scale study.
\newblock {\em Computer Networks}, 182:107495, 2020.
\newblock URL:
  \url{https://www.sciencedirect.com/science/article/pii/S1389128620311622},
  \href {https://doi.org/10.1016/j.comnet.2020.107495}
  {\path{doi:10.1016/j.comnet.2020.107495}}.

\bibitem{10.1145/2382016.2382018}
Juan~Camilo Cardona~Restrepo and Rade Stanojevic.
\newblock {{IXP}} traffic: A macroscopic view.
\newblock In {\em Proceedings of the 7th Latin American Networking Conference},
  Lanc '12, pages 1--8, New York, NY, USA, 2012. Association for Computing
  Machinery.
\newblock \href {https://doi.org/10.1145/2382016.2382018}
  {\path{doi:10.1145/2382016.2382018}}.

\bibitem{chatzisBenefitsUsingLarge2013}
Nikolaos Chatzis, Georgios Smaragdakis, Jan B{\"o}ttger, Thomas Krenc, and Anja
  Feldmann.
\newblock On the benefits of using a large {{IXP}} as an internet vantage
  point.
\newblock In {\em Proceedings of the 2013 Conference on {{Internet}}
  Measurement Conference}, {{IMC}} '13, pages 333--346, New York, NY, USA,
  October 2013. Association for Computing Machinery.
\newblock URL: \url{https://dl.acm.org/doi/10.1145/2504730.2504746}, \href
  {https://doi.org/10.1145/2504730.2504746}
  {\path{doi:10.1145/2504730.2504746}}.

\bibitem{chatzisThereMoreIXPs2013}
Nikolaos Chatzis, Georgios Smaragdakis, Anja Feldmann, and Walter Willinger.
\newblock There is more to {{IXPs}} than meets the eye.
\newblock {\em SIGCOMM Comput. Commun. Rev.}, 43(5):19--28, November 2013.
\newblock URL: \url{https://dl.acm.org/doi/10.1145/2541468.2541473}, \href
  {https://doi.org/10.1145/2541468.2541473}
  {\path{doi:10.1145/2541468.2541473}}.

\bibitem{cisco_air_2018_2023}
{Cisco Systems}.
\newblock Cisco annual internet report (2018--2023) white paper.
\newblock White {{Paper}} C11-741490, Cisco Systems, Inc., March 2020.
\newblock URL:
  \url{https://www.cisco.com/c/en/us/solutions/collateral/executive-perspectives/annual-internet-report/white-paper-c11-741490.pdf}.

\bibitem{10.1145/3419394.3423658}
Anja Feldmann, Oliver Gasser, Franziska Lichtblau, Enric Pujol, Ingmar Poese,
  Christoph Dietzel, Daniel Wagner, Matthias Wichtlhuber, Juan Tapiador, Narseo
  {Vallina-Rodriguez}, Oliver Hohlfeld, and Georgios Smaragdakis.
\newblock The lockdown effect: {{Implications}} of the {{COVID-19}} pandemic on
  internet traffic.
\newblock In {\em Proceedings of the {{ACM}} Internet Measurement Conference},
  Imc '20, pages 1--18, New York, NY, USA, 2020. Association for Computing
  Machinery.
\newblock \href {https://doi.org/10.1145/3419394.3423658}
  {\path{doi:10.1145/3419394.3423658}}.

\bibitem{10.1145/3452296.3472928}
Petros Gigis, Matt Calder, Lefteris Manassakis, George Nomikos, Vasileios
  Kotronis, Xenofontas Dimitropoulos, Ethan {Katz-Bassett}, and Georgios
  Smaragdakis.
\newblock Seven years in the life of {{Hypergiants}}' off-nets.
\newblock In {\em Proceedings of the 2021 {{ACM SIGCOMM}} 2021 Conference},
  Sigcomm '21, pages 516--533, New York, NY, USA, 2021. Association for
  Computing Machinery.
\newblock \href {https://doi.org/10.1145/3452296.3472928}
  {\path{doi:10.1145/3452296.3472928}}.

\bibitem{goldNetflixYouTubeAre2020}
Hadas Gold.
\newblock Netflix and {{YouTube}} are slowing down in {{Europe}} to keep the
  internet from breaking \textbar{} {{CNN Business}}, March 2020.
\newblock URL:
  \url{https://www.cnn.com/2020/03/19/tech/netflix-internet-overload-eu}.

\bibitem{grafana_12_0_0}
{Grafana Labs Contributors}.
\newblock Grafana/grafana: {{The}} open-source platform for monitoring and
  observability, May 2025.
\newblock URL: \url{https://github.com/grafana/grafana}.

\bibitem{10.1007/978-3-319-30505-9_25}
Samuel Henrique Bucke~Brito, Mateus A.~S.~Santos, R.~Fontes, D.~A.~L.~Perez,
  and Christian Esteve~Rothenberg.
\newblock Dissecting the largest national ecosystem of public internet
  {{eXchange}} points in brazil.
\newblock In {\em Passive and {{Active Network Measurement Conference}}}, pages
  333--345, Cham, 2016. Springer International Publishing.
\newblock URL:
  \url{https://www.semanticscholar.org/paper/4991ad5ceedfef67fc4be2b4d5fcae7f138ff59b},
  \href {https://doi.org/10.1007/978-3-319-30505-9_25}
  {\path{doi:10.1007/978-3-319-30505-9_25}}.

\bibitem{ixp_manager}
{INEX -- Internet Neutral Exchange Association}.
\newblock {{IXP}} manager --- world's most trusted {{IXP}} platform, 2025.
\newblock URL: \url{https://www.ixpmanager.org/}.

\bibitem{10.1145/1851182.1851194}
Craig Labovitz, Scott {Iekel-Johnson}, Danny McPherson, Jon Oberheide, and
  Farnam Jahanian.
\newblock Internet inter-domain traffic.
\newblock In {\em Proceedings of the {{ACM SIGCOMM}} 2010 Conference}, Sigcomm
  '10, pages 75--86, New York, NY, USA, 2010. Association for Computing
  Machinery.
\newblock \href {https://doi.org/10.1145/1851182.1851194}
  {\path{doi:10.1145/1851182.1851194}}.

\bibitem{10.1145/2602204.2602208}
Aemen Lodhi, Natalie Larson, Amogh Dhamdhere, Constantine Dovrolis, and
  kc~{claffy}.
\newblock Using {{peeringDB}} to understand the peering ecosystem.
\newblock {\em SIGCOMM Comput. Commun. Rev.}, 44(2):20--27, April 2014.
\newblock \href {https://doi.org/10.1145/2602204.2602208}
  {\path{doi:10.1145/2602204.2602208}}.

\bibitem{marwaha_gipr_2024}
Samir Marwaha.
\newblock Sandvine's 2024 global internet phenomena report: {{Global}} internet
  usage continues to grow, April 2024.
\newblock URL:
  \url{https://www.applogicnetworks.com/blog/sandvines-2024-global-internet-phenomena-report-global-internet-usage-continues-to-grow}.

\bibitem{openix_oix1_2016}
{Open-IX Association}.
\newblock {{IXP}} technical requirements --- {{OIX-1}} (version 2.0).
\newblock Standard OIX-1, Open-IX Association, October 2016.
\newblock URL:
  \url{https://storage.googleapis.com/website-v4_media/documents/IXPTechnicalRequirements2016.10.19.pdf}.

\bibitem{oix_about}
{Open-IX Association}.
\newblock About us -- open-{{IX}} association, 2025.
\newblock URL: \url{https://testing.oix.org/about/}.

\bibitem{PeeringDB}
{PeeringDB Community}.
\newblock {{PeeringDB}} --- the interconnection database, 2025.
\newblock URL: \url{https://www.peeringdb.com/}.

\bibitem{10.1145/3517745.3561462}
Maxime Piraux, Louis Navarre, Nicolas Rybowski, Olivier Bonaventure, and Benoit
  Donnet.
\newblock Revealing the evolution of a cloud provider through its network
  weather map.
\newblock In {\em Proceedings of the 22nd {{ACM}} Internet Measurement
  Conference}, Imc '22, pages 298--304, New York, NY, USA, 2022. Association
  for Computing Machinery.
\newblock \href {https://doi.org/10.1145/3517745.3561462}
  {\path{doi:10.1145/3517745.3561462}}.

\bibitem{richterPeeringPeeringsRole2014}
Philipp Richter, Georgios Smaragdakis, Anja Feldmann, Nikolaos Chatzis, Jan
  Boettger, and Walter Willinger.
\newblock Peering at {{Peerings}}: {{On}} the {{Role}} of {{IXP Route
  Servers}}.
\newblock In {\em Proceedings of the 2014 {{Conference}} on {{Internet
  Measurement Conference}}}, {{IMC}} '14, pages 31--44, New York, NY, USA,
  November 2014. Association for Computing Machinery.
\newblock URL: \url{https://dl.acm.org/doi/10.1145/2663716.2663757}, \href
  {https://doi.org/10.1145/2663716.2663757}
  {\path{doi:10.1145/2663716.2663757}}.

\bibitem{10.1145/637201.637213}
Matthew Roughan, Albert Greenberg, Charles Kalmanek, Michael Rumsewicz,
  Jennifer Yates, and Yin Zhang.
\newblock Experience in measuring backbone traffic variability: Models,
  metrics, measurements and meaning.
\newblock In {\em Proceedings of the 2nd {{ACM SIGCOMM}} Workshop on Internet
  Measurment}, Imw '02, pages 91--92, New York, NY, USA, 2002. Association for
  Computing Machinery.
\newblock \href {https://doi.org/10.1145/637201.637213}
  {\path{doi:10.1145/637201.637213}}.

\bibitem{sandvine_gipr_pr_2023}
{Sandvine}.
\newblock Sandvine's 2023 global internet phenomena report shows 24\% jump in
  video traffic, with netflix volume overtaking {{YouTube}}, January 2023.
\newblock URL:
  \url{https://www.applogicnetworks.com/press-releases/sandvines-2023-global-internet-phenomena-report-shows-24-jump-in-video-traffic-with-netflix-volume-overtaking-youtube}.

\bibitem{10.1145/3098822.3098853}
Brandon Schlinker, Hyojeong Kim, Timothy Cui, Ethan {Katz-Bassett}, Harsha~V.
  Madhyastha, Italo Cunha, James Quinn, Saif Hasan, Petr Lapukhov, and Hongyi
  Zeng.
\newblock Engineering egress with edge fabric: {{Steering}} oceans of content
  to the world.
\newblock In {\em Proceedings of the Conference of the {{ACM}} Special Interest
  Group on Data Communication}, Sigcomm '17, pages 418--431, New York, NY, USA,
  2017. Association for Computing Machinery.
\newblock \href {https://doi.org/10.1145/3098822.3098853}
  {\path{doi:10.1145/3098822.3098853}}.

\bibitem{tix_stats}
{Tanzania Internet eXchange (TIX)}.
\newblock {{TIX}} traffic statistics portal, 2025.
\newblock URL: \url{https://stats.tix.or.tz/}.

\bibitem{10.1007/978-3-031-56252-5_10}
Malte Tashiro, Romain Fontugne, and Kensuke Fukuda.
\newblock Following the~{{Data Trail}}: {{An Analysis}} of~{{IXP
  Dependencies}}.
\newblock In {\em Lecture Notes in Computer Science}, pages 199--227, Cham,
  January 2024. Springer Nature Switzerland.
\newblock \href {https://doi.org/10.1007/978-3-031-56252-5_10}
  {\path{doi:10.1007/978-3-031-56252-5_10}}.

\bibitem{tesseract_ocr_5_5_0}
{Tesseract OCR Contributors}.
\newblock Tesseract-ocr/tesseract: {{Tesseract}} open source {{OCR}} engine,
  November 2024.
\newblock URL: \url{https://github.com/tesseract-ocr/tesseract}.

\bibitem{timescaledb_2_19_3}
{TimescaleDB Contributors}.
\newblock Timescale/timescaledb: {{A}} time-series database for
  high-performance real-time analytics, April 2025.
\newblock URL: \url{https://github.com/timescale/timescaledb}.

\bibitem{tomeParis2024Olympics2024}
Jo{\~a}o Tom{\'e}.
\newblock Paris 2024 {{Olympics}} recap: {{Internet}} trends, cyber threats,
  and popular moments, August 2024.
\newblock URL: \url{https://blog.cloudflare.com/paris-2024-olympics-recap/}.

\bibitem{10.1145/3544216.3544268}
Matthias Wichtlhuber, Eric Strehle, Daniel Kopp, Lars Prepens, Stefan
  Stegmueller, Alina Rubina, Christoph Dietzel, and Oliver Hohlfeld.
\newblock {{IXP}} scrubber: Learning from blackholing traffic for {{ML-driven
  DDoS}} detection at scale.
\newblock In {\em Proceedings of the {{ACM SIGCOMM}} 2022 Conference}, Sigcomm
  '22, pages 707--722, New York, NY, USA, 2022. Association for Computing
  Machinery.
\newblock \href {https://doi.org/10.1145/3544216.3544268}
  {\path{doi:10.1145/3544216.3544268}}.

\bibitem{10.1145/3646547.3689679}
Semebia~Y. Wurah, Theophilus~A. Benson, and Edwin Mugume.
\newblock Poster: {{Analysis}} of the internet ecosystem in east africa:
  {{DNS}} resolution, {{ASN}} peering, and utilization of ixps.
\newblock In {\em Proceedings of the 2024 {{ACM}} on Internet Measurement
  Conference}, Imc '24, pages 785--786, New York, NY, USA, 2024. Association
  for Computing Machinery.
\newblock \href {https://doi.org/10.1145/3646547.3689679}
  {\path{doi:10.1145/3646547.3689679}}.

\bibitem{10.1145/3230543.3230544}
Tong Yang, Jie Jiang, Peng Liu, Qun Huang, Junzhi Gong, Yang Zhou, Rui Miao,
  Xiaoming Li, and Steve Uhlig.
\newblock Elastic sketch: Adaptive and fast network-wide measurements.
\newblock In {\em Proceedings of the 2018 Conference of the {{ACM}} Special
  Interest Group on Data Communication}, Sigcomm '18, pages 561--575, New York,
  NY, USA, 2018. Association for Computing Machinery.
\newblock \href {https://doi.org/10.1145/3230543.3230544}
  {\path{doi:10.1145/3230543.3230544}}.

\end{thebibliography}

\clearpage
\appendix
In \S\ref{ssec:representativeness-via-port-capacity}, we used port capacity to demonstrate that our dataset covers 87\% of total IXP infrastructure. A natural question follows: can port capacity---which PeeringDB reports even for IXPs that do not publish traffic statistics---serve as a proxy for traffic itself? If so, the growth patterns we observe may extend beyond the IXPs we directly measure. This appendix investigates that question by comparing capacity expansion across tracked and untracked IXPs, and by examining the stability of regional utilization over our two-year study period.

\section{Port Capacity as a Proxy for Traffic}\label{sec:supplementary}

We first examine the evolution of provisioned IXP port capacities~(\S\ref{subsec:capacity-expansion}) and show that capacity expansion is synchronized between tracked and untracked IXPs. We then compare these capacity growth trends with traffic growth~(\S\ref{subsec:utilization-stability}) and reveal how an increase in IXP port capacity is highly correlated with overall traffic growth, albeit with different (but steady) levels of utilization regionally. This tight coupling between infrastructure scaling and traffic demand supports applying our observed growth patterns to the broader IXP ecosystem.

\begin{figure}[t!]
	\begin{center}
		\begin{tikzpicture}
    \begin{axis}[
        tick align=outside,
        tick pos=left,
        axis line style=thick,
        axis x line*=bottom,
        x axis line style={},
        y axis line style={->, >=latex},
        axis y line*=left,
        major tick length=0.2cm,
        xtick style={color=black, thick},
        ytick style={color=black, thick},
        xticklabel style={font=\small},
        yticklabel style={font=\small},
        ylabel={Capacity (Tbps)},
        ymin=1000, ymax=2150,
        xmin=0, xmax=731,
        x tick label as interval=false,
        xtick={0, 90, 180, 270, 360, 450, 540, 630, 720},
        xticklabels={Jan, Apr, Jul, Oct, Jan, Apr, Jul, Oct, Dec},
        ytick={1000, 1300, 1600, 1900},
        yticklabels={1000, 1300, 1600, 1900},
        height=4.5cm,
        width=0.90\linewidth,
        clip=false,
        legend style={at={(0.5,1.05)}, anchor=center, draw=none, legend columns=2, column sep=0.1cm, row sep=0.05cm, font=\small, legend image post style={line width=2pt}, fill=none},
        legend cell align=left,
        set layers=standard
    ]
    
    % All data series
    \addplot[color={AsiaPacific}, thick] table[col sep=comma,x=day_index,y=all_actual] {figures/capacity/data.csv};
    \addlegendentry{Global}

    % Scraped data series
    \addplot[color={Africa}, thick] table[col sep=comma,x=day_index,y=scraped_actual] {figures/capacity/data.csv};
    \addlegendentry{Tracked}
    
    \end{axis}
\end{tikzpicture}
		\caption{Tracked (52\% growth) and total (54\% growth) provisioned port capacity trajectories align, reflecting ecosystem-wide scaling.}
		\label{fig:capacity}
	\end{center}
\end{figure}

\subsection{Capacity expansion}\label{subsec:capacity-expansion}

How is the aggregate capacity of IXPs increasing worldwide?
\cref{fig:capacity} compares the growth trajectories of all IXPs listed in PeeringDB as of January 2023 against our 472 tracked IXPs.
Over our two-year study, total provisioned capacity in PeeringDB increased by 54\%, from 1125 Tbps to 1910 Tbps.
In parallel, our 472 tracked IXPs showed nearly identical growth of 52\%, from 1140 Tbps to 1700 Tbps.

This synchronization implies that infrastructure scaling across the IXP ecosystem is driven uniformly by common factors such as anticipated demand, technological advancement, and competitive market pressures, further validating port capacity as a proxy for traffic growth.

While aggregate port capacity increases smoothly and linearly, individual IXP port capacity grows in steps, as expected.
\cref{fig:ports} plots weekly port capacity updates for SwissIX and Namex Rome.
For these IXPs, weekly port capacity increases range from 10 Gbps to 300 Gbps.
These updates are not always monotonic, as IXPs occasionally report slight reductions in port capacity.
However, overall, IXPs typically consistently increase their port capacities.

\begin{figure}[t!]
	\begin{center}
		\begin{tikzpicture}
    \begin{axis}[
        tick align=outside,
        tick pos=left,
        axis line style=thick,
        axis x line*=bottom,
        x axis line style={},
        y axis line style={->, >=latex},
        axis y line*=left,
        major tick length=0.2cm,
        xtick style={color=black, thick},
        ytick style={color=black, thick},
        xticklabel style={font=\small},
        yticklabel style={font=\small},
        ylabel={Capacity (Gbps)},
        ymin=2500, ymax=7000,
        xmin=0, xmax=731,
        xtick={0, 90, 180, 270, 360, 450, 540, 630, 720},
        xticklabels={Jan, Apr, Jul, Oct, Jan, Apr, Jul, Oct, Dec},
        ytick={2500, 3500, 4500, 5500, 6500, 7500},
        yticklabels={2500, 3500, 4500, 5500, 6500, 7500},
        height=4.5cm,
        width=0.90\linewidth,
        clip=false,
        legend style={at={(0.5,1.05)}, anchor=center, draw=none, legend columns=2, column sep=0.1cm, row sep=0.05cm, font=\small, legend image post style={line width=2pt}, fill=none},
        legend cell align=left,
        set layers=standard
    ]

    \addplot[color={Europe}, thick] table[col sep=comma,x=day,y=capacity] {figures/ports/data/swiss_ix.csv};
    \addlegendentry{SwissIX}
    
    \addplot[color=black, thick] table[col sep=comma,x=day,y=capacity] {figures/ports/data/namex_rome.csv};
    \addlegendentry{Namex Rome}
    
    \end{axis}
\end{tikzpicture}
		\caption{Individual IXP port capacity (SwissIX, Namex Rome) grows in discrete steps.}
		\label{fig:ports}
	\end{center}
\end{figure}

\begin{figure}[t!]
	\begin{center}
		\begin{tikzpicture}
    \begin{axis}[
        tick align=outside,
        tick pos=left,
        axis line style=thick,
        axis x line*=bottom,
        x axis line style={},
        y axis line style={->, >=latex},
        axis y line*=left,
        major tick length=0.2cm,
        xtick style={color=black, thick},
        ytick style={color=black, thick},
        xticklabel style={font=\small},
        yticklabel style={font=\small},
        ylabel={Utilization (\%)},
        ymin=0, ymax=17,
        xmin=0, xmax=731,
        xtick={0, 90, 180, 270, 360, 450, 540, 630, 720},
        xticklabels={Jan, Apr, Jul, Oct, Jan, Apr, Jul, Oct, Dec},
        ytick={0, 2, 4, 6, 8, 10, 12, 14},
        yticklabels={0, 2, 4, 6, 8, 10, 12, 14},
        height=4.5cm,
        width=0.90\linewidth,
        clip=false,
        legend style={at={(0.42,1.25)}, anchor=center, draw=none, legend columns=4, column sep=0.1cm, row sep=0.05cm, font=\small, legend image post style={line width=2pt}, fill=none},
        legend cell align=left,
        set layers=standard
    ]
    
    % Total utilization (solid line)
    \addplot[color={Africa}, thick] table[col sep=comma,x=day_index,y=africa] {figures/utility/regional_data.csv};
    \addlegendentry{Africa}

    \addplot[color={AsiaPacific}, thick] table[col sep=comma,x=day_index,y=asia_pacific] {figures/utility/regional_data.csv};
    \addlegendentry{Asia-Pacific}

    \addplot[color={Australia}, thick] table[col sep=comma,x=day_index,y=australia] {figures/utility/regional_data.csv};
    \addlegendentry{Australia}

    \addplot[color={Europe}, thick] table[col sep=comma,x=day_index,y=europe] {figures/utility/regional_data.csv};
    \addlegendentry{Europe}
 
    \addplot[color={MiddleEast}, thick] table[col sep=comma,x=day_index,y=middle_east] {figures/utility/regional_data.csv};
    \addlegendentry{Middle East}
    
    \addplot[color={NorthAmerica}, thick] table[col sep=comma,x=day_index,y=north_america] {figures/utility/regional_data.csv};
    \addlegendentry{North America}
    
    \addplot[color={SouthAmerica}, thick] table[col sep=comma,x=day_index,y=south_america] {figures/utility/regional_data.csv};
    \addlegendentry{South America}

    \addplot[color={black}, thick] table[col sep=comma,x=day_index,y=utility] {figures/utility/total_data.csv};
    \addlegendentry{Global}

    \end{axis}
\end{tikzpicture}
		\caption{Regional utilization varies but remains stable over two years despite 49\% traffic growth.}
        \label{fig:utilization}
	\end{center}
\end{figure}

\subsection{Regional utilization}\label{subsec:utilization-stability}

To assess whether user traffic drives IXP capacity provisioning, this subsection examines regional utilization of our 472 monitored IXPs.
Regional network utilization is defined as the ratio of total average daily traffic observed from our 472 monitored IXPs in a region to their total provisioned public peering port capacity, based on contemporaneous PeeringDB data.
As depicted in \cref{fig:utilization}, we find striking stability in regional utilization rates over our two-year study.
For example, Africa maintains exceptionally consistent utilization, averaging 8.75\% with minimal overall fluctuation ($\pm$0.18\%).
Similarly, Australia (mean 3.57\%, -0.41\% change) and North America (mean 4.96\%, +0.86\% change) also demonstrate strong consistency.

Some regions show slightly more pronounced trends, but changes remain modest.
Europe's utilization declined from 9.27\% to 7.05\% (-2.22\%), while South America and the Middle East showed decreases of -1.19\% and -2.60\%, respectively.
The Asia-Pacific region uniquely experienced an increase, rising from 7.99\% to 10.71\% (+2.72\%).

Despite significant traffic growth in our dataset (\S\ref{subsec:global-growth-trajectory-and-volatility}), stable utilization patterns highlight consistent provisioning strategies by IXPs and their peering networks.
This stability suggests a deliberate operational approach to maintaining predictable capacity headroom relative to observed demand.
Although average utilization setpoints vary significantly by region, reflecting local economic conditions, peering density, and engineering practices, their consistent intra-regional stability points to a high degree of operational predictability within the IXP landscape, reinforcing port capacity as a proxy for traffic dynamics.

Aggregate utilization of our 472 monitored IXPs remained stable during the two-year measurement period.
Overall, our 472 tracked IXPs experienced a 52\% capacity increase, closely paralleled by a 49\% rise in aggregate observed daily traffic.
This proportionality underscores a direct, robust relationship between physical infrastructure investment and realized traffic growth within our extensive sample.
Such tightly coupled scaling confirms a foundational aspect of the public IXP ecosystem: infrastructure provisioning closely anticipates and matches traffic demand, reflecting a well-coordinated approach to capacity management across the IXP community.
Thus, although we cannot measure their traffic, the port capacity of missed IXPs may provide insights into their overall traffic volumes.

For example, if our dataset---covering 87\% of total IXP port capacity---observes approximately 300 Tbps of peak traffic, extrapolation suggests an additional 45 Tbps traversing the unmonitored 13\% of capacity.
Future work should refine this capacity-based estimation by developing regional utilization benchmarks.
The stable relationship between infrastructure and usage offers a powerful tool for illuminating the broader Internet ecosystem, and the dataset presented in this paper provides a foundation for such analysis.

\end{document}